\documentclass[floatfix,twocolumn,nofootinbib,superscriptaddress,a4paper,preprintnumbers]{revtex4-1}

\usepackage{amsmath}
\usepackage{amssymb}
\usepackage{graphicx}
\usepackage{ulem}
\usepackage{mathrsfs}
\usepackage{color}
\usepackage[T1]{fontenc} 
\usepackage{slashed}
\usepackage{xspace}
\usepackage{hyperref}
\usepackage{xspace}
\usepackage{soul}
\usepackage{booktabs}

\usepackage{multirow}

\newcommand{\MS}{{\ensuremath{\overline{\text{MS}}}}\xspace}
\newcommand{\os}{\text{OS}}

\newcommand\SARAH{{\tt SARAH}\xspace}
\newcommand\SPheno{{\tt SPheno}\xspace}

\newcommand{\nn}{\nonumber}
\newcommand{\ztssm}{$\mathbb{Z}_2$SSM~}

\DeclareMathOperator{\llog}{\overline{\text{log}}}

 \definecolor{mkgreen}{rgb}{0.2,.70,.3}

\begin{document}

\preprint{ BONN-TH-2017-11 \ \
 KA-TP-37-2017 \ \ MITP/17-089}

\title{$N$-loop running should be combined with $N$-loop matching}

\author{Johannes~Braathen} 
\email{braathen@lpthe.jussieu.fr}
\affiliation{LPTHE, UPMC Univ.~Paris 6, Sorbonne Universit\'es, 4 Place Jussieu, F-75252 Paris, France}
\affiliation{LPTHE, CNRS, 4 Place Jussieu, F-75252 Paris, France }

\author{Mark~D.~Goodsell} 
\email{goodsell@lpthe.jussieu.fr}
\affiliation{LPTHE, UPMC Univ.~Paris 6, Sorbonne Universit\'es, 4 Place Jussieu, F-75252 Paris, France}
\affiliation{LPTHE, CNRS, 4 Place Jussieu, F-75252 Paris, France }

\author{Manuel E. Krauss} 
\email{mkrauss@th.physik.uni-bonn.de}
\affiliation{Bethe Center for Theoretical Physics \& Physikalisches Institut der 
Universit\"at Bonn,\\ Nu{\ss}allee 12, 53115 Bonn, Germany}

\author{Toby Opferkuch} 
\email{toopferk@uni-mainz.de} 
\affiliation{Bethe Center for Theoretical Physics \& Physikalisches Institut der 
Universit\"at Bonn,\\ Nu{\ss}allee 12, 53115 Bonn, Germany}
\affiliation{PRISMA Cluster of Excellence \& Mainz Institute for Theoretical Physics,\\
Johannes Gutenberg-Universit\"at Mainz, 55099 Mainz, Germany}

\author{Florian Staub}
\email{florian.staub@kit.edu}
\affiliation{Institute for Theoretical Physics (ITP), Karlsruhe Institute of Technology, Engesserstra{\ss}e 7, D-76128 Karlsruhe, Germany}
\affiliation{Institute for Nuclear Physics (IKP), Karlsruhe Institute of Technology, Hermann-von-Helmholtz-Platz 1, D-76344 Eggenstein-Leopoldshafen, Germany}


\begin{abstract}
We investigate the high-scale behaviour of Higgs sectors beyond the Standard Model, pointing out that the proper matching of the quartic couplings before applying the renormalisation group equations (RGEs) is of crucial importance for reliable predictions at larger energy scales. In particular, the common practice of leading-order parameters in the RGE evolution is insufficient to make precise statements on a given model's UV behaviour, typically resulting in uncertainties of many orders of magnitude. We argue that, before applying $N$-loop RGEs, a matching should even be performed at $N$-loop order in contrast to common lore. We show both analytical and numerical results where the impact is sizeable for three minimal extensions of the Standard Model: a singlet extension, a second Higgs doublet and finally vector-like quarks. We highlight that the known two-loop RGEs tend to moderate the running of their one-loop counterparts, typically delaying the appearance of Landau poles. For the addition of vector-like quarks we show that the complete two-loop matching and RGE evolution hints at a stabilisation of the electroweak vacuum at high energies, in contrast to results in the literature.

\end{abstract}
\maketitle

\section{Introduction}
Minimal extensions of the Standard Model (SM) are invaluable tools in the pursuit of physics beyond the SM (BSM). They offer the possibility of studying different effects at energy scales testable by the Large Hadron Collider (LHC) in a comparably clean environment -- i.e. the models typically contain the minimal numbers of new fields to exhibit novel phenomenology. All additional states also alter the high-scale behaviour of the model compared to the SM
expectations. For instance, it is known today that the SM becomes metastable if it is extrapolated to very high energies \cite{Buttazzo:2013uya,Degrassi:2012ry,Andreassen:2017rzq,Spencer-Smith:2014woa}: 
at a scale of $10^{9-11}$ GeV the quartic coupling runs negative. The scale at which the potential becomes unstable could be significantly affected 
by the presence of new states -- it could even completely disappear. This would then indicate that the BSM model is valid up to the Planck scale. An opposite effect can occur if large couplings are present. In this case, a Landau pole might be present which points towards the breakdown of the 
theory. Both effects, the presence of Landau poles or deeper vacua, can also be used to directly constrain the parameters of a new physics model. 
A parameter point can be discarded if the model becomes strongly interacting at energies already probed by the LHC, or if the life-time of the electroweak breaking vacuum is too short 
on cosmological time scales. 

Many of these effects have already been studied in the literature for plethora of different models such as singlet extensions \cite{Lerner:2009xg,Lebedev:2012zw, EliasMiro:2012ay, Pruna:2013bma,Costa:2014qga}, triplet extensions \cite{Hamada:2015bra,Khan:2016sxm}, Two-Higgs-Doublet-Models (THDMs)  \cite{Cheon:2012rh,Chakrabarty:2014aya,Chakrabarty:2016smc,Ferreira:2015rha,Chakrabarty:2017qkh,Chowdhury:2015yja,
Gori:2017qwg,Basler:2017nzu,Dev:2014yca,Das:2015mwa}
or models with vector-like fermions \cite{Xiao:2014kba}. These studies utilise the one- and sometimes even two-loop renormalisation group equations (RGEs). However, less
care was was taken in the determination of the parameters which enter the RGE running. Often, two-loop RGEs were combined with a \emph{tree-level} matching.

A proper determination including higher-order corrections of the quartic coupling, which enters the RGE running, was so far only performed for the SM \cite{Buttazzo:2013uya}.\footnote{Loop corrections in the scalar sector were taken into account in Ref.~\cite{Kanemura:2015fra} for a singlet extension
and in Refs.~\cite{Basler:2016obg,Basler:2017uxn} for a THDM. These studies did not however investigate the impact on the high-scale behaviour of the model.
In Ref.~\cite{Kobakhidze:2013pya} in turn, a one-loop matching has been performed for that purpose in the context of a seesaw-II as well as a left-right symmetric model.} 
It was shown that even the next-to-next-to-leading-order (NNLO) shifts to $\lambda$ are important for determining the fate of the model. 
This is remarkable, because it is well known that the loop corrections to the Higgs mass are small if they are calculated at $Q=m_t$: the corresponding shifts in $\lambda$ are only 2.5\%. While 
the corrections from top quarks are of a similar order in many BSM models, other corrections like the ones from Higgs self-interactions can be much larger. This has been recently pointed out in 
Ref.~\cite{Braathen:2017izn} in the example of THDMs where the two-loop corrections to the Higgs masses were calculated for the first time.\footnote{In this context, it was pointed out that using ``on-shell'' masses and couplings as input can 
be quite dangerous because it hides the presence of large couplings which could even spoil perturbation theory. A similar observation was made for another model, the Georgi-Machacek model, 
in Ref.~\cite{Krauss:2017xpj}.} In this context it is important to note that the common lore ``$N$-loop running needs $N-1$-loop matching'' is applicable only in certain scenarios. This degree of matching catches the leading logarithms correctly but misses finite contributions which can be relevant in many BSM applications.

We show in this work that higher order corrections can be very important for the study of the UV behaviour of a theory leading to four main conclusions:
\begin{enumerate}
\item The threshold corrections at low energy can lead to substantial (finite) shifts in the running parameters of a model: therefore $N$-loop RGEs with $N$-loop matching is required for consistency.
\item The change from one-loop to two-loop running can flatten the running at large values of the coupling, preventing the onset of a Landau pole at high energies -- leading to a form of asymptotic safety.
\item Alternatively, in the case where the running drives some quartic coupling negative, higher-order corrections can lead to significant changes to the predicted scale of metastability.
\item As a by-product of the above, we find that new \emph{fermionic} fields at low energies can stabilise the SM potential.
\end{enumerate}
We illustrate the above with a detailed examination of three examples: a singlet extension, the SM extended by vector-like quarks and the THDM.

The paper is organised at follows. In Sec.~\ref{sec:matching} we give a step-by-step prescription for the general matching procedure including effects at the $N-1$- and $N$-loop levels. In Sec.~\ref{sec:setup} we give details into the 
procedure used to obtain higher order corrections to the quartic couplings in the different models considered, before we discuss in Sec.~\ref{sec:results} the numerical results, providing insights including approximate formulae.

\section{Matching and Running}
\label{sec:matching}

To extrapolate a theory from the electroweak scale to high energies, we require two ingredients:
\begin{enumerate}
 \item The value of the couplings at the ``low scale''  where the running starts.
 \item The RGE running of all parameters.
\end{enumerate}

\subsection{Renormalisation Group Equations}
We shall always work in the \MS scheme. In this scheme, the $\beta$-functions, which describe the energy dependence 
of the parameters $\Theta$, 
are defined as
\begin{equation}
\beta_i = \mu \frac{d \Theta_i}{d\mu} \,.
\end{equation}
Here, $\mu$ is an arbitrary mass scale. $\beta_i$ can be expanded in a perturbative series:
\begin{equation}
\beta_i = \sum_n \frac{1}{(16 \pi^2)^n} \beta_i^{(n)} 
\end{equation}
$\beta_i^{(1)}$ and $\beta_i^{(2)}$ are the one- and two-loop contributions  to the running which we are mainly interested in.
The expressions for the two-loop running of the parameters appearing in a given model can be obtained from the 
generic expressions valid for a general quantum field theory, given in Refs.~\cite{Machacek:1983tz,Machacek:1983fi,Machacek:1984zw,Luo:2002ti}.

\subsection{Matching}
\label{sec:Matching}
The renormalised coupling constants $\Theta_i$ in $d=4-2\epsilon$ dimensions, which enter the running, are
related to the corresponding bare couplings $\Theta_i^0$ by
\begin{equation}
\Theta_i^0 \mu^{-C_i \epsilon} = \Theta_i + \sum_n \frac{a_i^{(n)}}{\epsilon^n}\,.
\end{equation}
Here,  $C_i$ are constant factors depending on the character of $\Theta_i$\footnote{Gauge and Yukawa couplings 
have $C_i=1$, quartic couplings $C_i=2$.}. The coefficients $a_i$ are the result of a perturbative expansion. 
In general, two approaches are possible to determine the \MS parameters as function of physical observables
such as masses. 
\begin{enumerate}
 \item In an on-shell calculation the physical observables are identical at each loop-level, but all finite and infinite 
 corrections are absorbed into the counter-terms of the Lagrangian parameters ($\delta \Theta_i^{\os}$). 
 \item In an \MS calculation the counter-terms of the Lagrangian parameters (${\delta  \Theta}_i^{\MS}$)
 contain only the divergences. Therefore, the 
calculated masses depend on the loop-level at which the calculation is performed. 
\end{enumerate}
The bare Lagrangian parameters are identical in both cases
\begin{equation}
\Theta^0_i = \Theta_i^{\os} - \delta \Theta_i^{\os} = 
\Theta_i(\mu)-{\delta  \Theta}_i^{\MS} \,.
\end{equation}
In an on-shell calculation, however, there is no generic set of renormalisation group equations known, and therefore to explore a theory at high energies it is necessary to use \MS equations -- i.e. to extract the underlying \MS parameters of the theory, and then run them. On the other hand, in an \MS calculation, the physical parameters are functions of the Lagrangian parameters: so if we are given the physical quantities, we must invert these functions to extract the \MS ones. This is where complications appear, and why many studies in the literature resort to simply using tree-level matching. 

For example, suppose that we want to extract the quartic coupling of the SM from the Higgs mass. The Higgs mass $M_h$ is, however, calculated in terms of the underlying Lagrangian parameters as a loop expansion via the on-shell condition
\begin{align}
M_h^2 = \lambda v^2 + \sum_{n=1}^\infty \frac{1}{(16\pi^2)^n} \Delta^{(n)} M_h^2 (\lambda)\,.
\label{eq:generalmh}\end{align}
This is in general a highly non-linear equation in $\lambda$; but fortunately since the series is perturbative we can solve it through expanding 
\begin{equation}
 \lambda=\lambda^{(0)}+\frac{1}{16\pi^2}\delta^{(1)}\lambda+\frac{1}{(16\pi^2)^2}\delta^{(2)}\lambda +\dotso
\end{equation}
 to find
\begin{align}
\lambda^{(0)} &= \frac{M_h^2}{ v^2}\,,\nn\\
\delta^{(1)}\lambda&=-\frac{1}{v^2}\Delta^{(1)}M_h^2 \left. \right|_{\lambda =\lambda^{(0)}}\,, \\
\delta^{(2)}\lambda&=-\frac{1}{v^2}\bigg[\delta^{(1)}\lambda\frac{\partial}{\partial\lambda}\Delta^{(1)} M_h^2  +\Delta^{(2)}M_h^2 \bigg]_{\lambda =\lambda^{(0)}}\,,\nn
\label{eq:perturbativesolveforlambda}\end{align}
which is simple enough for the Standard Model and extensions without scalar mixing -- so we shall give analytical expressions in Secs.~\ref{sec:Z2Analytic} and \ref{sec:VL}. On the other hand, for more complicated models, we need to solve Eq.~(\ref{eq:generalmh}) through iteration, and we shall adopt this approach in general for the numerical studies. Whichever way we solve for $\lambda$, as we shall argue in the next subsection, when we are using $N$-loop RGEs, to obtain a consistent expansion of our \MS parameters in terms of the physical ones we should use $N$-loop matching.

\subsection{Loop Level of Matching and Running}
It is commonly accepted in matching a high-energy theory onto an effective field theory that, if the running is performed using $N$-loop RGEs, it is only necessary to use $N-1$-loop threshold corrections. However, the rationale for this criterion is less well known: it corresponds to matching the \emph{logarithmic terms} of the calculations, and neglects the finite parts. To take perhaps the simplest and best-known example, suppose that we want to integrate out two heavy Dirac fermions with masses $M_2>  M_1$ that couple with charges $Q_2, Q_1$ to some $U(1)$ gauge theory. Suppose that the contribution of other fields to the one-loop beta function is $b_0$, so at high energies the beta function is $b_0 + \frac{4}{3}(Q_1^2 + Q_2^2)$, and we take the gauge coupling at high energies to be $g(\Lambda)$. Now, to determine the gauge coupling at a low energy $\mu$ the classic prescription is to run first to $M_2$, match at tree level, then run to $M_1$, match again, and then run down to $\mu$. This gives the one-loop value of the gauge coupling to be 
\begin{align}
\frac{ 8\pi^2}{g^2(\mu)} =& \frac{8\pi^2 }{g^2 (\Lambda)} - (b_0 + \frac{4}{3} Q_1^2 + \frac{4}{3} Q_2^2) \log \frac{M_2}{\Lambda} \nn\\
&- (b_0 + \frac{4}{3} Q_1^2) \log \frac{M_1}{M_2} - b_0 \log \frac{\mu}{M_1}\,.
\end{align}
This can be rewritten as 
\begin{align*}
\frac{ 8\pi^2}{g^2(\mu)} &=  \frac{8\pi^2 }{g^2 (\Lambda)}  - b_0 \log \frac{\mu}{\Lambda} - \frac{4}{3} Q_1^2 \log \frac{M_1}{\Lambda}- \frac{4}{3} Q_2^2 \log \frac{M_2}{\Lambda}\,,
\end{align*}
which is also the result of simply matching the two theories at the scale $\mu$ and including the threshold corrections: for corrections to the gauge coupling, the thresholds contain only logarithmic corrections. However, suppose that instead we wanted to match the two theories at a scale $M$ that we have chosen to be different to $M_1$ and/or $M_2$ -- for example, because other sectors of the theory contain particles of that mass. Now, if we just match the two theories at tree level, we would find after first running to $M$ and then matching:
\begin{align}
\frac{ 8\pi^2}{\tilde{g}^2(\mu)} &= \frac{8\pi^2 }{g^2 (\Lambda)} - (b_0 + \frac{4}{3} Q_1^2 + \frac{4}{3} Q_2^2) \log \frac{M}{\Lambda} - b_0 \log \frac{\mu}{M} \nn\\
&= \frac{ 8\pi^2}{g^2(\mu)} - \frac{4}{3} Q_1^2  \log \frac{M_1}{M}- \frac{4}{3} Q_2^2  \log \frac{M_2}{M}
\end{align}
In this case, we have a difference between the two values of the gauge coupling \emph{at one loop}: in other words, we conclude that for single-scale matching, even in this simple case there is a discrepancy of one-loop order between the two calculations. What we have captured is, as mentioned above, the \emph{logarithmic terms only}.

The reader might complain that this is a slightly strange example since the threshold corrections themselves contain only logarithmic terms. We can instead take another example, more relevant for this work, of the quartic coupling of a real scalar $\phi$ coupled to a Majorana (Weyl) fermion $\psi$:
\begin{align}
\mathcal{L} \supset - \frac{1}{24} \lambda \phi^4 - \bigg(\frac{1}{2} (m + y \phi) \psi \psi + h.c. \bigg). 
\end{align}
If $y$ and $m$ are real, then the one-loop threshold correction to $\lambda$ from integrating out the fermion $\psi$ at matching scale $M$ is
\begin{align}
\delta \lambda =& -\frac{ y^2}{16\pi^2} \bigg[ \mathbf{ 32 y^2} + (24 y^2 - 4 \lambda)\log \frac{m}{M} \bigg].
\end{align}
The finite, non-logarithmic correction is highlighted in bold. On the other hand, 
\begin{align}
\frac{d\lambda}{d \log \mu} =& -\frac{1}{16\pi^2} \bigg[ y^2( 24 y^2 - 4  \lambda) - 3 \lambda^2 \bigg]\,. \label{EQ:SimpleRGE}
\end{align}
Hence the tree-level matching clearly finds the correct logarithmic terms, but we miss the finite correction of $-\frac{2y^2}{\pi^2} $, even at one loop. 

One argument against using $N$-loop matching has been that the logarithmic corrections should be the most important: if the couplings are all small, then e.g. $\frac{2y^2}{\pi^2}$ should be a very small correction and we will only have significant contributions to the couplings when we run down from high energies from the terms enhanced by logarithms. However, this preconception is biased from the idea of the SM where the strong gauge coupling and top Yukawa run to smaller values at high energies, and the quartic runs to zero; this also applies for models with no new scalars such as Split Supersymmetry, so for example we see that the supersymmetric threshold corrections become smaller as we take the supersymmetry scale higher -- when the theory in the ultraviolet is the Minimal Supersymmetric SM. 

For a generic BSM theory -- especially ones with new quartic scalar couplings -- this expectation is no longer true. Indeed, if we neglect other iterations then it is well known that a quartic coupling will tend to \emph{increase} as we increase the energy -- see e.g. Eq.~(\ref{EQ:SimpleRGE}) with $y=0$. For a generic theory with new quartics, as we increase their values relative to the Yukawa or gauge couplings we expect for some given value there will be a transition from decreasing couplings in the UV (leading to metastability or instability of the potential) to the appearance of a Landau pole. In this latter case, since the quartic couplings will be even larger at high energies, including the finite parts of threshold corrections becomes \emph{vital} for a consistent matching. 

So far we have discussed threshold corrections when integrating out a heavy theory at high energies. This should not be confused with the matching that we need to do when we are running \emph{up} to investigate the appearance of a Landau pole or the vacuum stability of a model. When we are applying threshold corrections at low energies (around the electroweak scale), then if we want to investigate whether a theory is stable, metastable or has a cut-off, and if so at what scale, then clearly it is important that the starting point for our calculation is determined accurately. For example, taking our toy model above and neglecting the coupling $y$, then integrating up from $M$ to a scale $\mu$ we find a Landau pole at approximately
\begin{align}
\Lambda \approx M \exp \left( - \frac{16\pi^2}{3 \lambda(M)} \right)\,,
\end{align}
but if we have incorrectly determined $\lambda(M)$ by an amount $\delta \lambda$, then the ratio of the correct cut-off scale to the erroneous one $\Lambda^\prime $ is
\begin{align}
\frac{\Lambda^\prime }{\Lambda} \simeq& \exp \left(\frac{16\pi^2}{3} \frac{\delta \lambda}{\lambda^2 (M)} \right)\,, \nn\\
\simeq&\, 200 \exp \left(\frac{16\pi^2}{3} \left(\frac{\delta \lambda}{ \lambda^2(M)}-0.1\right) \right)\,.
\end{align}
In the Standard Model, the difference between the tree-level value for $\lambda$ and the two-loop value is tiny when the extraction is performed at the top mass. However, as we shall see, in other models a shift of $10\%$ in the quartic coupling is conservative (and we should not forget the famous example of the MSSM where $\delta \lambda \gtrsim \lambda$).

\section{Numerical set-up}
\label{sec:setup}
For the numerical calculations we make use of the {\tt Mathematica} package \SARAH \cite{Staub:2008uz,Staub:2009bi,Staub:2010jh,Staub:2012pb,Staub:2013tta,Staub:2015kfa} 
to produce a spectrum generator based on \SPheno \cite{Porod:2003um,Porod:2011nf,Staub:2017jnp}. \SPheno includes routines to obtain the full one-loop corrections to all masses 
as well as the two-loop corrections to real scalars. 
The two-loop calculations are done in the gauge-less limit and based on the generic results of Refs.~\cite{Martin:2003it,Goodsell:2015ira}. In non-SUSY BSM models, these results suffer in general
from the so called ``Goldstone boson catastrophe'' even in 
the gauge-less limit because the couplings of the Goldstone do not disappear in this limit, but are proportional to the cubic and quartic potential parameters. Therefore, we also make use of the 
results of Ref.~\cite{Braathen:2016cqe} which provides a general solution to this problem. \\

In practice, we perform the following steps to calculate the mass spectrum based on a set of \MS parameters:

\begin{enumerate}
 \item The running couplings $\Theta(Q)$ at the scale $Q=m_t$ are taken as input, while the SM parameters are evolved to this scale including all known SM corrections, i.e. three-loop running and two-loop matching for strong coupling $g_3$ and top Yukawa $Y_t$.  
 \item The tree-level tadpoles $T_i$ are solved to fix the remaining free parameters, which in what follows are typically the mass parameters $\mu_i^2 |\phi|^2$. 
 \item The tree-level mass are calculated by diagonalising the tree-level mass matrices
 \item The $n$-loop corrections to the tadpoles $\delta^{(n)} t_i$ are calculated. The imposed renormalisation conditions are
 \begin{equation}
  T_i + \sum_j^n \delta^{(j)} t_i = 0\,, 
 \end{equation}
 which cause shifts in $\mu_i^2$:
 \begin{equation}
 \label{eq:shiftsm2}
  \mu_i^2 \to \mu_i^2 + \sum_j^n \delta^{(j)} \mu_i^2\,. 
 \end{equation}
\item The one- and two-loop self-energies for real scalars are calculated for external gauge eigenstates.  
The pole masses are the eigenvalues of the loop-corrected mass matrix calculated as
 \begin{equation}
  M_{\phi}^{(n)}(p^2) = \tilde{M}^{(2L)}_{\phi} - \sum_j^n \Pi_{\phi}^{(j)}(p^2) \,.
 \end{equation}
Here, $\tilde{M}_{\phi}$ is the tree-level mass matrix including the shifts (\ref{eq:shiftsm2}). \\
The calculation of the one-loop self-energies in both cases is done iteratively for each eigenvalue $i$ until the on-shell condition
\begin{equation}
  \left[\text{eig} M_{\phi}^{(n)}(p^2=m^2_{\phi_i})\right]_i \equiv m_{\phi_i}^2
\end{equation}
is fulfilled. The renormalised rotation matrix is taken to be the one calculated for $p^2=m_{\phi_1}^2$.
\end{enumerate}

If a chosen set of input parameters $\Theta(Q)$ results in the desired physical masses and mixing angles when using a $N$-loop calculation, we refer to them  
as $N$-loop couplings. Thus, with tree-level relations we have leading-order (LO) parameters, while the one- and two-loop mass corrections result in the next-to-leading-order (NLO) and NNLO couplings, respectively.

Finding the correct set of \MS couplings corresponding to the desired physical parameters at loop level is non-trivial. In what follows we use a simple fitting algorithm which varies the input parameters until the desired masses and mixing angles are obtained.


\section{Models and results}
\label{sec:results}

\subsection{Singlet Extension}
\label{SEC:singletext}
\begin{figure}[tb]
\includegraphics[width=1.\linewidth]{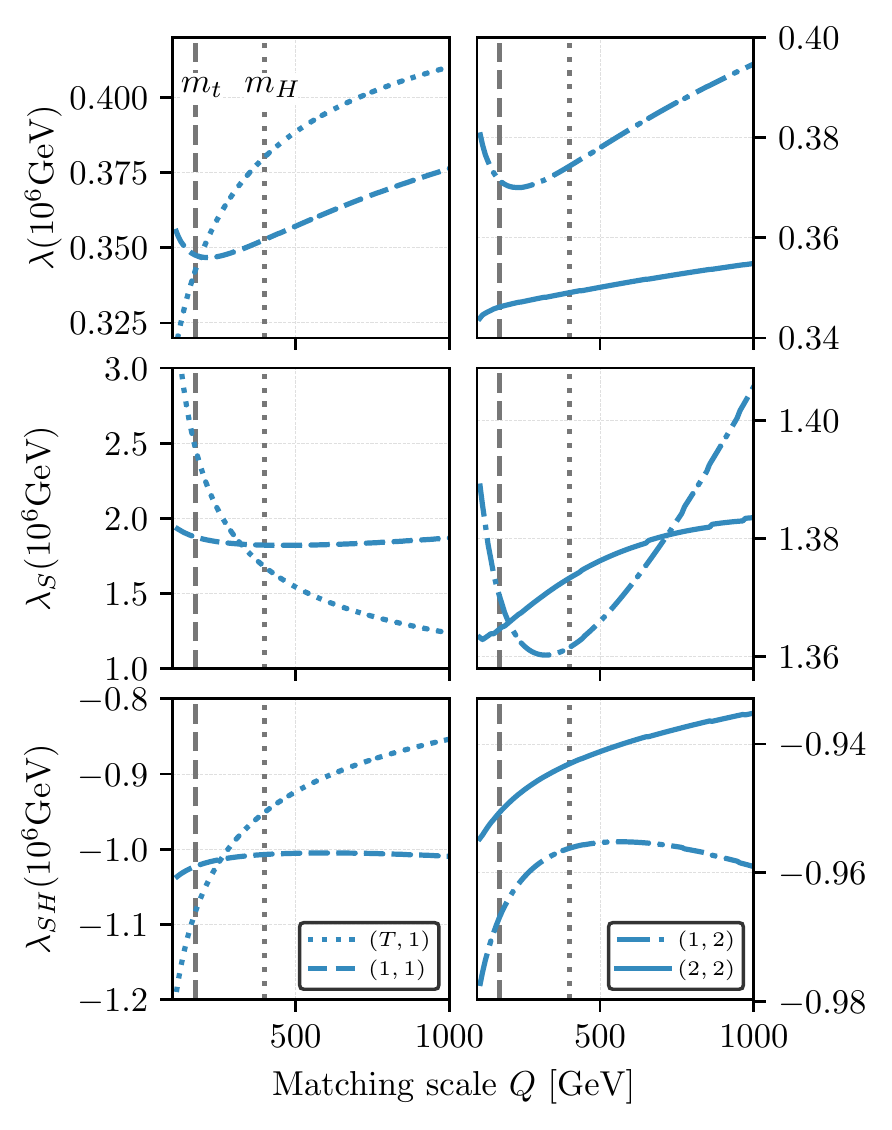}
\caption{Values of the running quartic couplings at the scale $Q=10^6$~GeV using one-loop (left) and two-loop RGEs (right) as functions of the matching scale at which the quartic couplings were calculated. The labels $(n,m)$ refer to $n$-loop level matching of the quartics and $m$-loop RGEs. We choose the parameters of the singlet extended SM at the matching scale to be $m_h=125$~GeV, $m_H=400$~GeV, $\tan\alpha=0.3$ and $v_S=300$~GeV. Cubic terms were set to zero to ensure a scale invariant input. Note that the $y$-axes ranges are different in each panel.}
\label{fig:Q}
\end{figure}
We start with the SM extended by a real gauge singlet $S$. The potential  
reads
\begin{align}
V &= \mu^2 |H|^2 + \frac12 M_S^2 S^2 + \kappa_1 |H|^2 S + \frac13 \kappa_2 S^3\nonumber \\
 &\qquad+\frac12 \lambda |H|^4 + \frac12 \lambda_{SH} S^2 |H|^2 + \frac12 \lambda_S S^4\,.
\end{align}
After electroweak symmetry breaking the CP-even scalar components mix to two physical states $h$, $H$ via a rotation angle $\alpha$. At tree level we can use 
$m_h$, $m_H$ and $t_\alpha \equiv\tan\alpha$ as input to calculate the quartic couplings 
\begin{align}\label{eq:SSM_LO_quartics_1}
\lambda_H = & \frac{m_h^2 + m_H^2 t_\alpha^2}{v^2(1+t_\alpha^2)}\,, \\
\lambda_S = & \frac{\kappa_1  v^2}{8 v_S^3}- \frac{\kappa_2}{4 v_S}+\frac{(m_H^2 +  m_h^2 t_\alpha^2 )}{4 (1 + t_\alpha^2) v_S^2}\,, \label{eq:SSM_LO_quartics_2}\\
\lambda_{SH} = & -\frac{m_H^2 t_\alpha - m_h^2 t_\alpha + \kappa_1 v + \kappa_1 t_\alpha^2 v}{v v_S (1+ t_\alpha^2) }\,,\label{eq:SSM_LO_quartics_3}
\end{align}
where $v$ and $v_S$ are the vacuum expectation values (VEVs) of the Higgs doublet and singlet scalar respectively. Interpreting $m_h$, $m_H$ and $t_\alpha$ as physical on-shell parameters, the quartic calculated via Eqs.~(\ref{eq:SSM_LO_quartics_1})-(\ref{eq:SSM_LO_quartics_3}) are therefore the LO values. 

Aside from the typical requirement that the quartic couplings remain perturbative, the constraints from perturbative unitarity need to be taken into account. For that, we can evaluate the scalar $2\to 2$ scattering amplitude in the limit of high energies and demand that the eigenvalues stay below $8\,\pi$. The 
conditions for the model under consideration read
 \begin{align}
 |\lambda_{SH}| < 8 \pi\,, \\
 |\lambda| < 8 \pi\,, \\
 \left|-6 \lambda_S - 3 \lambda \pm \sqrt{4 \lambda_{SH}^2 + 9 (\lambda - 2 \lambda_S)^2}\right| < 16 \pi\,, \label{eq:SSM_unitarity}
 \end{align}
which leads to $\lambda_S < 4\pi/3$ for small $\lambda$ and $\lambda_{SH}$. Although a seemingly weak constraint at first sight, this can become a severe constraint particularly in the case of small $v_S$, cf. Eq.~(\ref{eq:SSM_LO_quartics_2}).

\begin{table}[tb]
\begin{tabular}{cccccc}
\toprule
 $(n,m)$ & $\lambda$ & $\lambda_S$ & $\lambda_{SH}$ & $\Lambda_{4\pi}$  [GeV] & $\Lambda_{4\pi}^{\rm{unit.}}$  [GeV] \\\midrule 
$(T,1)$ &  \multirow{2}{*}{0.34}&\multirow{2}{*}{1.1} & \multirow{2}{*}{-1.1}            & $6.4\cdot 10^3$ & $3.2\cdot 10^3$ \\
$(T,2)$ &                      & &              & $8.0\cdot 10^6$ & $1.3\cdot 10^4$\\
\midrule
$(1,1)$ &  \multirow{2}{*}{0.33}&\multirow{2}{*}{0.24} & \multirow{2}{*}{-0.97}             & $6.4\cdot 10^8$ & $3.2\cdot 10^8$ \\
$(1,2)$ &                      &&               &  $1.3\cdot  10^{12}$ & $2.5\cdot 10^9$ \\
\midrule
$(2,1)$ &  \multirow{2}{*}{0.32}&\multirow{2}{*}{0.18} & \multirow{2}{*}{-0.94}             & $5.1\cdot  10^{10}$ & $2.5\cdot 10^{10}$ \\
$(2,2)$ &                       &&              & $1.0\cdot 10^{14}$ & $2.0\cdot 10^{11}$\\
\bottomrule
\end{tabular}
\caption{
Values of the quartic couplings and the cut-off for different combinations of parameters, $\lambda^{(n)}$, and RGEs, $\beta^{(m)}$ at the loop orders $n$ and $m$ respectively. $\Lambda_{4\pi}$ is the scale at which the quartic couplings first exceed $4\pi$ while $\Lambda_{4\pi}^{\rm{unit.}}$ is the naive $4\pi$ cut-off augmented by the unitarity constraint of Eq.~(\ref{eq:SSM_unitarity}). The observables are fixed at the $n$-loop order to be $m_h=125$~GeV, $m_H=700$~GeV, $\tan\alpha=0.1$ while the remaining input parameters are chosen as $\kappa_1=0$~GeV, $\kappa_2=2000$~GeV, $v_S=175$~GeV.}
\label{tab:SSMBP2}
\end{table}
\begin{figure}[tb]
\includegraphics[width=\linewidth]{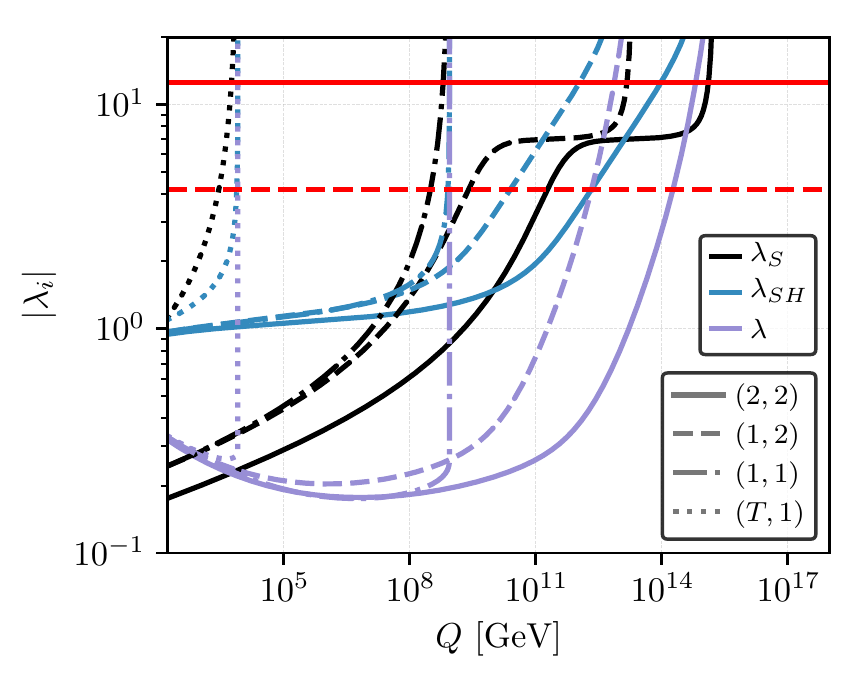}
\caption{The running of the quartic couplings for the point given in Tab.~\ref{tab:SSMBP2}. The line styles refer to the loop order of the matching and RGE running as described in Tab.~\ref{tab:SSMBP2}, namely $(n,m)$ refers to the matching at $n$-loop order with $m$-loop RGEs. The solid red line is the $4\pi$ perturbativity limit, while the dashed red line is the unitarity constraint of $4\pi/3$ obtained from Eq.~(\ref{eq:SSM_unitarity}) in the limit $\lambda_S \gg \lambda_{SH},\lambda$.}
\label{fig:SSM:running}
\end{figure}

In the following, we compare different approaches for the matching of the quartic couplings. In Fig.~\ref{fig:Q}, we show the values of the quartic couplings at the scale $10^6\,$GeV as a function of the scale $Q$ where the matching is performed. This is done by first running the SM RGEs to the scale $Q$ where we then match the quartic couplings to the spectrum at tree level, one loop as well as two loop. The final step is the running of the singlet extended SM RGEs (both at both one- or two-loop order) up to $10^6\,$GeV. In the left-hand plots, we use one-loop RGEs for the cases of tree-level matching (dotted) and one-loop matching (dashed). Although there is no dependence of the quartics on the matching scale when using tree-level matching, the scale dependence induced by the RGEs is larger than for the case of one-loop matching and one-loop RGEs. In the right-hand panels, we show the quartics at $10^6\,$GeV, evaluated with two-loop RGEs, when using one-loop (dashed) and two-loop matching (solid lines). Once again, the scale dependence is decreased when using $N$-loop matching with $N$-loop RGEs, where in this case $N=2$. 

Also visible in Fig.~\ref{fig:Q} are the large differences between the eventual coupling values when using the ``traditional'' approach of tree-level matching with one-loop RGEs and the approach we advertise, two-loop matching and two-loop RGEs. This also means that large differences are expected when evaluating the cut-off scale of a theory, i.e. the scale at which the model becomes non-perturbative or violates unitarity. In Tab.~\ref{tab:SSMBP2}, we show the cut-off scale of a particular 
parameter point when using $n$-loop matching in conjunction with $m$-loop RGEs. 
We show the scale at which the quartics become non-perturbative ($\Lambda_{4\pi}$) separately to the case where either perturbativity or unitarity is violated ($\Lambda_{4\pi}^{\rm{unit.}}$). The corresponding running of the individual couplings is displayed in Fig.~\ref{fig:SSM:running}. Note that we only display the cases $(n,m)=(T,1),\, (1,1), \,(1,2)$ and $(2,2)$ in this figure as dotted, dot-dashed, dashed and full lines corresponding to the comparison of $N$ versus $N-1$ matching.
The impact of the two-loop RGEs is a moderation of the one-loop RGEs: while the one-loop $\beta$-function of $\lambda_S$ is given by $\beta^{(1)}_{\lambda_S} = \frac{1}{16\pi^2} (36 \lambda_S^2 + \lambda_{SH}^2)$, so that $\lambda_S$ tends to grow very rapidly, there is a moderating term from the two-loop RGEs which goes with $\frac{1}{(16\pi^2)^2} (-816 \lambda_S^3 - 20 \lambda_S \lambda_{SH}^2 )$. Therefore, using the one-loop RGEs only,  $\lambda_S$ grows large very quickly -- whereas the unitarity limit is reached at a much later scale when using two-loop RGEs. 
Nevertheless, a complete stalling of the evolution is typically only reached at $\lambda_S$ values which already violate the unitarity limit according to Eq.~(\ref{eq:SSM_unitarity}), see the black dashed (full) line between $10^9$ and $10^{13}\,$GeV ($10^{12}$ and $10^{15}\,$GeV).
The moderation of the evolution of $\lambda_{SH}$  and $\lambda$ is not as pronounced. For $\lambda_{SH}$, the corresponding $\beta$-function grows with $\frac{1}{16\pi^2} 12 \lambda_S \lambda_{SH}$ with only a small moderating effect from the two-loop RGEs.
As a consequence, it becomes larger than $4\,\pi$ before $\lambda_S$ and then drags the latter with it.
In total, in particular because of the large two-loop contributions to $\beta_{\lambda_S}$, there can be several orders of magnitude between the eventual cut-off scales when using one- or two-loop RGEs.

The effect of using a two-loop matching instead of a tree matching, in turn, is a reduction of the quartic couplings. The reason is the positive mass corrections to $m_H$, leading to smaller \MS couplings when doing the proper loop-level matching.
As shown here, the impact can be large and we observe positive shifts in the eventual cut-off scale by several orders of magnitude when including the matching.

\begin{figure*}[tb]
\includegraphics{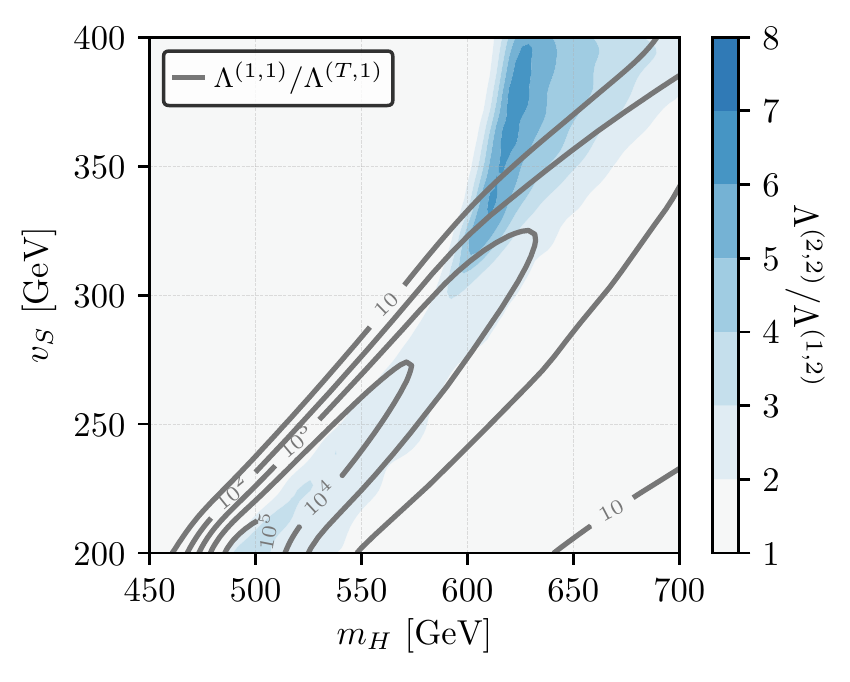} \hspace{0.1cm} \includegraphics{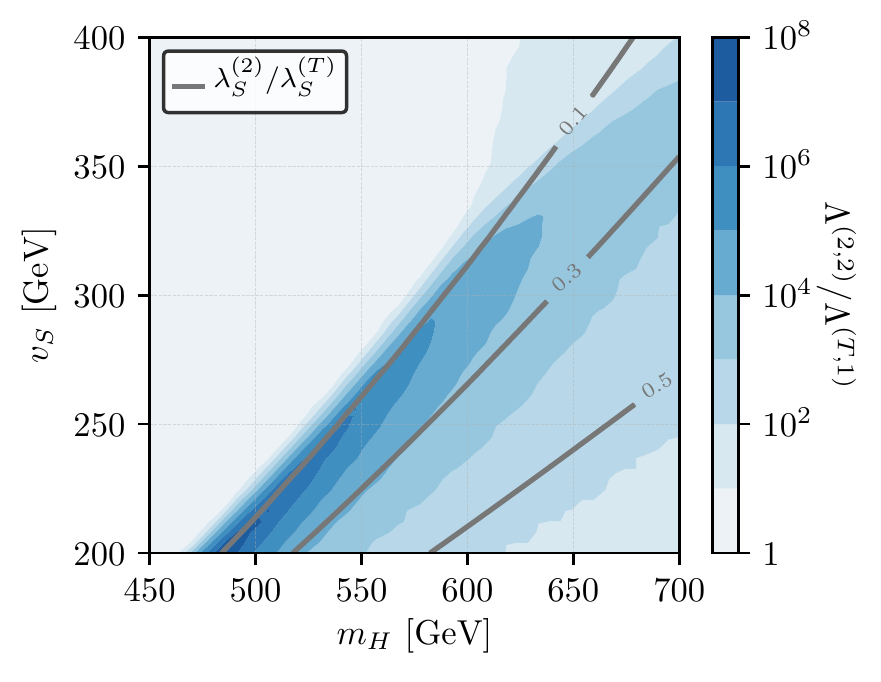}
\caption{Difference in the predicted cut-off scale depending on the matching performed as a function of the singlet VEV $v_S$ and the heavy CP-even Higgs mass $m_H$. {\it Left:} We show the ratio of the obtained cut-off given matching at $N$ versus $N-1$ order using the RGEs at $N$-loop order. The coloured(grey) contours use the two(one)-loop RGEs, therefore showing the ratio of the matching at two (one) loop versus one loop (tree level), respectively. 
{\it Right:} Ratio of the most precise calculation performed (both matching and RGEs at two-loop order) versus the leading order (tree-level matching and one-loop RGEs). The grey contours correspond to the ratios of the quartic coupling $\lambda_S$ for these two scenarios. Here we have fixed the physical parameters such that $m_h=125$~GeV, $\tan\alpha=0.2$, while the remaining parameters are chosen as $\kappa_1=0$~GeV and $\kappa_2=1000$~GeV.}
\label{fig:SSM_mH_vS_DeltaLam}
\end{figure*}

Finally in Fig.~\ref{fig:SSM_mH_vS_DeltaLam} we show in the $m_H$-$v_S$ plane the differences between using $N-1$-loop and $N$-loop matching when applying $N$-loop RGE running. 
The cut-off scale here and in what follows is defined as the scale at which \emph{either} one of the couplings grows larger than $4\pi$ \emph{or} any of the conditions for perturbative unitarity are violated, each evaluated with the running \MS  quartic couplings. The grey contours in the left-hand pane of Fig.~\ref{fig:SSM_mH_vS_DeltaLam} display the ratio of the evaluated cut-off scales for $N=1$. In particular for small $v_S$, which leads to large quartic couplings, the effects are quite drastic as loop effects become very important. The differences between one- and two-loop matching (shown as blue coloured contours) are significantly milder in this region, the maximum difference is just a factor of three. For large $v_S$, instead, the quartic couplings are comparably small, leading to large cut-off scales in general. This also means, however, that during the long RGE running, small shifts in couplings can lead to more drastic effects as is seen in the upper region in the plot with $v_S \gtrsim 350\,$GeV. However, the cut-off scale differences stay below an order of magnitude for $N=2$.

On the right-hand side of Fig.~\ref{fig:SSM_mH_vS_DeltaLam} we present 
the difference in cut-off scales between the most extreme cases, 
tree-level matching using one-loop RGEs versus two-loop matching using two-loop RGEs. 
In particular for small values of the singlet VEV, the eventual cut-off scale can be many orders of magnitude 
larger than the cut-off scale evaluated with tree-level matching. Grey contours show the ratios of the singlet
quartic couplings  at $Q=m_t$ between the two matching approaches,
$\lambda_S^{(2)}/\lambda_S^{(T)}$. Already at the matching scale, differences of an order of magnitude between tree and two-loop matching 
can appear, emphasising the requirement for proper matching and running when analysing the high-scale behaviour of a given model.

\subsection{Singlet Extension with an additional $\mathbb{Z}_2$ symmetry}
\subsubsection{Analytical approximation}
\label{sec:Z2Analytic}
We may now make a further simplification to the singlet extension studied in Sec.~\ref{SEC:singletext}, namely adding an additional $\mathbb{Z}_2$ symmetry under which the singlet scalar is charged -- for clarity we will call this model the $\mathbb{Z}_2$SSM to distinguish it from the SSM. This symmetry forbids non-zero values for the couplings $\kappa_1$, $\kappa_2$ and for the singlet VEV $v_S$, and furthermore eliminates mixing in the Higgs sector. Therefore, the derivation of analytic expressions for the radiative corrections to the matching of the Higgs quartic coupling $\lambda$, and their comparison to numerical studies, are significantly simpler, and follow the procedure outlined in Sec.~\ref{sec:Matching}. Here we will be interested in the part of the corrections that come on top of the purely SM corrections due to the singlet scalar\, and shall give expressions including two-loop contributions. 

The one- and two-loop corrections to the Higgs mass in the SM are well-known and small; however, in our model there may be large corrections from the singlet scalar.
%
In order to extract the two-loop contributions via Eq.~(\ref{eq:perturbativesolveforlambda}) we require the two-loop mass correction, and also the derivatives of the one-loop part. However, our two-loop calculation is performed in the gaugless limit in Feynman gauge, so we require the full one-loop Higgs mass correction in this limit:
\begin{align}
\Delta^{(1)} M_h^2 (p^2)&= 3 y_t^2 (4m_t^2-p^2) B(m_t^2,m_t^2)- \frac{3}{2} \lambda^2 v^2 B(0,0) \nn\\
 - \frac{9}{2}  \lambda^2 v^2 & B(m_h^2,m_h^2)- \frac{1}{2} v^2 \lambda_{SH}^2 B(m_S^2,m_S^2).
\end{align}
Here we have defined $m_S^2 \equiv M_S^2 + \frac{1}{2} \lambda_{SH} v^2$, $m_h^2 \equiv \lambda v^2$ which are the tree-level squared masses of the singlet and Higgs respectively, while a complete list of the definitions for our loop functions can be found in Refs.~\cite{Braathen:2016cqe,Braathen:2017izn} which are based upon the basis defined in Refs.~\cite{Martin:2003qz,Martin:2003it}. This gives us
\begin{align}
\lambda^{(0)} =& \,\frac{M_h^2}{ v^2} \nn\\
\delta^{(1)}\lambda=&\,\delta^{(1)}_{\rm SM} \lambda + \frac{1}{2} \lambda_{SH}^2 B(m_S^2,m_S^2) \nn\\
\delta^{(2)}\lambda=&\,\delta^{(2)}_{\rm SM} \lambda -\frac{1}{v^2}\bigg[\Delta^{(2)}_{\rm \mathbb{Z}_2 SSM} M_h^2  \bigg]_{\lambda =\lambda^{(0)}} \nn\\
 + \frac{3}{2} \lambda_{SH}^2 &\lambda B(m_S^2, m_S^2) \bigg( 3 B (m_h^2, m_h^2) \nn\\
&+ 3 m_h^2 B((m_h^2)',m_h^2)  + B(0,0) \bigg).
\end{align}
where we use the shorthand notation $B(x',y)\equiv\frac{\partial}{\partial x}B(x,y)$. We note that the infra-red-divergent piece $B(0,0)$ will cancel against an equivalent piece from $\Delta^{(2)}_{\rm \mathbb{Z}_2 SSM} M_h^2  $, similarly to the effect noted in Ref.~\cite{Martin:2014cxa}.

For the \ztssm, we obtain
\begin{align}
 \delta^{(2)}\lambda=&\,\delta^{(2)}_{\rm SM} \lambda \nn\\-\frac{1}{2} \lambda_{SH} \bigg(&\lambda_{SH}^3  v^2  M_{SSSSS}(m_S^2,m_S^2,m_S^2,m_S^2,m_h^2)\nn\\
 &+6 \lambda    \lambda_{SH}^2 v^2   M_{SSSSS}(m_h^2,m_S^2,m_h^2,   m_S^2,m_S^2)\nn\\
 &-6 \lambda  \lambda_{SH}  U_{SSSS}(m_h^2,m_h^2,m_S^2,m_S^2)\nn\\
 &-4 \lambda_{SH}^2  U_{SSSS}(m_S^2,m_S^2,m_h^2   ,m_S^2)\nn\\
 &+9 \lambda ^2\lambda_{SH} v^2   V_{SSSSS}(m_h^2,m_h^2,m_h^2,m_S^2,m_S^2)\nn\\
 &+2 \lambda_{SH}^3  v^2  V_{SSSSS}(m_S^2,m_S^2,m_S^2,m_h^2,m_S^2)\nn\\
 &-\lambda_{SH}^2    Y_{SSSS}(m_S^2,m_S^2,m_S^2,m_h^2)\nn\\
 &-9 \lambda ^2    Y_{SSSS}(m_h^2,m_h^2,m_h^2,m_S^2)\nn\\
 &-12 \lambda_{SH}\lambda _S     Y_{SSSS}(m_S^2,m_S^2,m_S^2,m_S^2)\nn\\
 &-6 \lambda_{SH}   \lambda _S    Z_{SSSS}(m_S^2,m_S^2,m_S^2,m_S^2)\nn\\
 &-12 \lambda  \lambda_{SH}    Z_{SSSS}(m_h^2,m_h^2,m_S^2,m_S^2)\bigg) \nn\\
  - \frac{1}{2v^2}\lambda_{SH}^2 &\bigg( S_{SSS} (m_h^2, m_S^2, m_S^2) - I (m_h^2, m_S^2, m_S^2)\bigg)\nn \\
  &\hspace{-44pt}+ \frac{9}{2} \lambda_{SH}^2 \lambda m_h^2 B (m_S^2, m_S^2) B((m_h^2)',m_h^2).
\end{align}
This expression is valid for the gaugeless limit but with generic external momentum (so we can take the momentum in the loop integrals on-shell as the procedure demands, if we wish). However, if we take the ``generalised effective potential limit'' introduced in Refs.~\cite{Braathen:2016cqe,Braathen:2017izn} and employed in \SARAH, then the penultimate line vanishes and the loop functions simplify considerably. We can then obtain a further simplified version of this expression by replacing $m_h^2$ by its tree-level value $\lambda v^2$ and by performing an expansion in powers of $v^2/m_S^2$ and keeping only the leading and sub-leading terms, giving
\begin{align}
 \delta^{(2)}\lambda\simeq \delta^{(2)}_{\rm SM} \lambda& -\frac{9}{4v^2}\lambda_{SH}\lambda A(m_S^2)\\
 &+\lambda_{SH}^3\big[1-2\llog m_S^2+\llog^2 m_S^2\big]\nn\\
 &+\frac{1}{4}\lambda_{SH}^2\lambda\big[-18-6\llog^2 m_S^2\nn\\
 &\hspace{45pt}+(36\llog m_h^2-12)\llog m_S^2\big]\nn\\
 &+3\lambda_{SH}^2\lambda_S\big[-1+\llog m_S^2+\llog^2 m_S^2\big]\,. \nn
\end{align} 
\subsubsection{Numerical study}
\begin{figure*}[tb]
\includegraphics{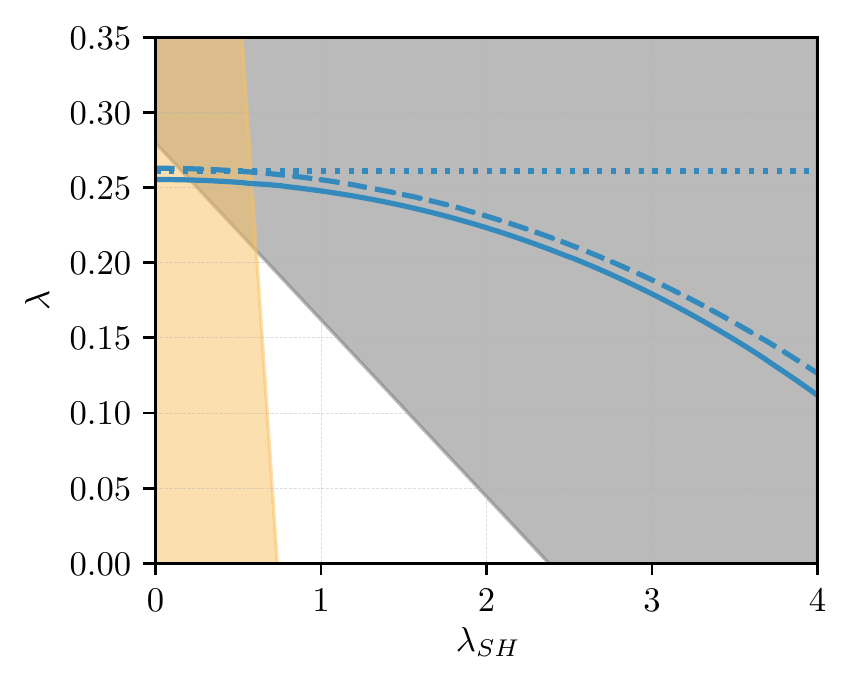}\hspace{0.1cm}\includegraphics{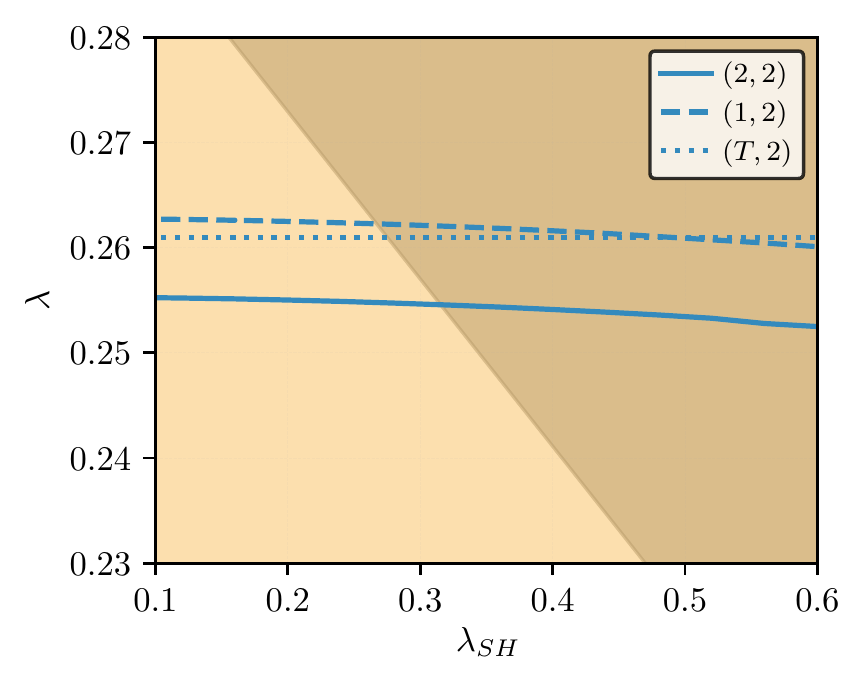}
\caption{Different phases in the \ztssm shown in the $\lambda_{SH}-\lambda$ plane, where the couplings are taken at the scale $Q=m_t$. The orange shaded region of parameter space corresponds to UV-complete theories, $i.e.$ none of the three quartic couplings ($\lambda$, $\lambda_{SH}$, $\lambda_S$) become non-perturbative and the constraints from unitarity are not violated before the Planck scale; the black shaded region corresponds to theories with stable vacua. The thin blue lines give $\lambda$ as a function of $\lambda_{SH}$ when imposing $m_h=125.15\text{ GeV}$ with a matching condition at respectively tree-level (dotted curve), one-loop order (dashed curve) and two-loop order (solid line). The other parameters of the scalar sector are $\lambda_S=0.1$, and $M_S=500\text{ GeV}$.}
\label{fig:Z2SSM_phases}
\end{figure*}

Because the $\mathbb{Z}_2$ symmetry forbids some couplings, the corrections to the matching conditions can be understood in terms of only three parameters added to the SM ones: $\lambda_{SH}$, $M_S$, and (to a lesser extent) $\lambda_S$. The effects of using loop-corrected matching and RGEs in the \ztssm are similar to those observed in Sec.~\ref{SEC:singletext} for the SSM, although for most values of $\lambda_{SH}$ and $M_S$ the shift to the quartic coupling has only a very small effect on the value of the cut-off scale. We give in Tab.~\ref{tab:Z2SSMlambdas} our results for $\lambda$ obtained for the three different orders of matching, for both small and large $\lambda_{SH}$ and for two choices of $M_S$. For small $\lambda_{SH}$ the one-loop shift to $\lambda$ is small, because of a cancellation between the purely-SM part -- dominated by the effect of the top quark -- and the singlet part of $\delta^{(1)}\lambda$. If one then considers larger values of $\lambda_{SH}$, the term from the singlet becomes dominant over the SM one, and $\delta^{(1)}\lambda$ is a large negative shift -- the evolution of $\lambda$, extracted at different orders, as a function of $\lambda_{SH}$ is also shown in Fig.~\ref{fig:Z2SSM_phases}, discussed below. At two loops however, there is no cancellation between SM and singlet contributions, and $\delta^{(2)}\lambda$ is always a negative shift to the Higgs quartic, as was observed previously for the general SSM. On the other hand, it is always small, showing -- importantly -- that perturbativity of the model is preserved.
\begin{table}[tb]
\begin{tabular}{ccccc}
\toprule
 $M_S$ [GeV] & $\lambda_{SH}$ & $\lambda_\text{tree}$ & $\lambda_{1\ell}$ & $\lambda_{2\ell}$\\
\midrule 
  \multirow{2}{*}{500} & 0.25 & 0.2610 & 0.2623 & 0.2551 \\
 & 3 & 0.2610 & 0.1885 & 0.1794 \\
\midrule
\multirow{2}{*}{1000} & 0.25 & 0.2610 & 0.2651 & 0.2546 \\
 & 3 & 0.2610 & 0.1548 & 0.1385 \\
\bottomrule
\end{tabular}
\caption{Values of the Higgs quartic $\lambda$ obtained from matchings at tree-level, one-loop and two-loop orders, for different choices of $M_S$ and $\lambda_{SH}$. The singlet quartic coupling $\lambda_S$ is set to be 0.1. }
 \label{tab:Z2SSMlambdas}
\end{table}

Having fewer parameters allows for a more detailed study of the different phases of the theory. Indeed, there are two transitions that occur respectively:
\begin{itemize}
 \item Between a metastable and a stable vacuum of the theory; for the physically relevant values of $\lambda$ around 0.25-0.26, this happens for $\lambda_{SH}\sim 0.3$ and depends very little on $M_S$ or $\lambda_S$. 
 \item Between a UV-complete model and a UV-incomplete one -- in other words the cut-off scale of the model becomes smaller than the Planck scale for sufficiently large couplings.
\end{itemize}

\begin{figure}[tb]
\includegraphics[width=\linewidth]{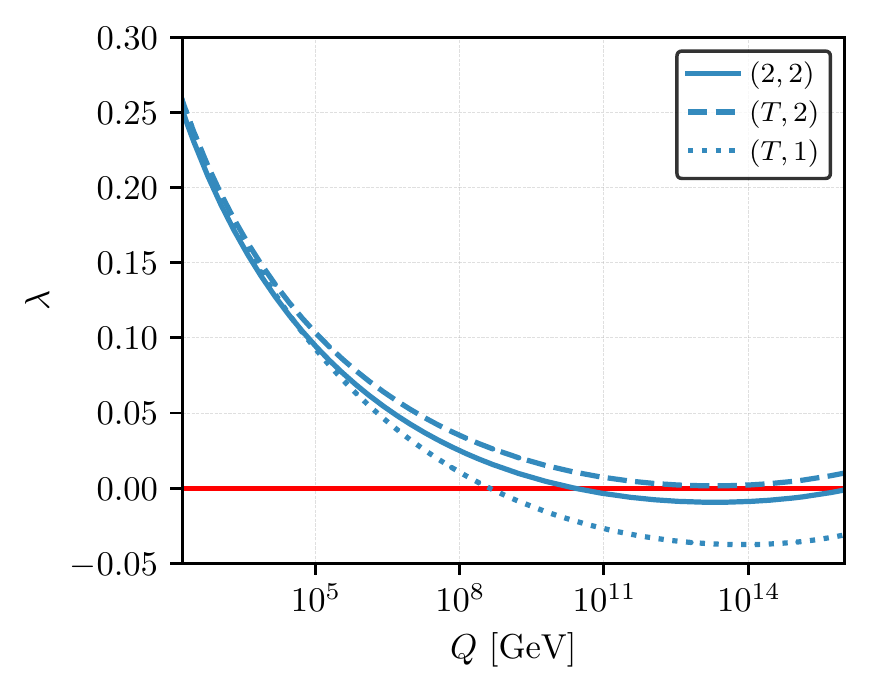}
\caption{Running of the Higgs quartic coupling as a function of the renormalisation scale $Q$, having taken $\lambda_{SH}=0.28$, $\lambda_S=0.1$ and $M_S=500\text{ GeV}$. The value of $\lambda(m_t)$ is obtained by requiring $m_h=125.15\text{ GeV}$, with different orders of matchings depending on the curve. The solid line corresponds to the use of two-loop matching and two-loop RGE running, the dashed line to tree-level matching and two-loop RGEs, and the dotted line to tree-level matching and one-loop RGEs. Note that because of the cancellation that occurs in the one-loop correction for small $\lambda_{SH}$ (discussed in the main text), the curves we would have obtained using one-loop matching would have been very similar to those with tree-level matching.}
\label{fig:Z2SSM_lambda_run}
\end{figure}

Fig.~\ref{fig:Z2SSM_lambda_run} shows an example in which the order of the matching performed to extract $\lambda$ causes differences in the stability of the vacuum of the theory. Indeed, while the curve with two-loop matching and two-loop RGE running (solid line) crosses to negative values of $\lambda$ -- for $10^{10}\text{ GeV}\lesssim Q\lesssim 10^{16}\text{ GeV}$ -- the curve with tree-level matching (dashed line) does not, because of the negative shift to the initial value of $\lambda$ at scale $Q=m_t$ at two-loop order. Two-loop corrections to the matching of $\lambda$ may exclude some parameter points that appear viable when only using a tree-level matching and are therefore important in the discussion of allowed regions of parameter space. Comparing the dashed and dotted lines, we also observe the stabilising effect of the use of the two-loop RGEs, as discussion in Sec.~\ref{SEC:singletext}.

Fig.~\ref{fig:Z2SSM_phases} shows how both types of transitions occur in this model. The different domains in this figure were obtained as follows:    we start with values of the couplings, at the scale $Q=m_t$, in the range $\lambda\in[0,0.35]$ and $\lambda_{SH}\in[0,4]$, and take $\lambda_S=0.1$ and $M_S=500\text{ GeV}$. We then use two-loop RGEs to run the couplings up to the Planck scale, and we verify whether the Higgs quartic $\lambda$ becomes negative at any point, and whether perturbativity or unitarity are lost below the Planck scale. The left panel of Fig.~\ref{fig:Z2SSM_phases} presents the whole range of couplings that we considered, while the right panel shows an enlargement of the region in which the transition between stable and metastable phases occurs.  

We observe that the UV-complete phase of the model corresponds to smaller values of the inputs at scale $Q=m_t$ -- which can easily be understood as large values of the couplings at $m_t$ naturally lead to even larger values at higher scales. Furthermore, we can see that the phase of the model with stable vacua is associated with larger values of $\lambda_{SH}$, and that when $\lambda$ decreases, the value of $\lambda_{SH}$ needed to ensure a stable vacuum increases. While the SM part of the $\beta$-function of $\lambda$ is negative and tends to drive it to negative values, the additional piece in $\beta_\lambda$ in the \ztssm is positive and is of the form $\beta_\lambda^{(1)}\supset\frac{1}{16\pi^2}\lambda_{SH}^2$. When lowering $\lambda(m_t)$ a higher value of $\lambda_{SH}$ is needed so that the $\beta$-function of $\lambda$ changes sign earlier, and that $\lambda$ does not run negative at some scale.

The blue lines in Fig.~\ref{fig:Z2SSM_phases} give $\lambda(m_t)$ obtained from requiring that $m_h=125.1\text{ GeV}$ as a function of $\lambda_{SH}$. The different curves correspond to the different orders at which the matching can be done: dotted for tree-level matching, dashed for one-loop and solid for two-loop order. The most important point to notice is that, as for the vacuum stability (see Fig.~\ref{fig:Z2SSM_lambda_run}), there is a value of $\lambda_{SH}$ -- here around 0.65 -- for which the UV-completeness -- in other words whether perturbativity or unitarity is broken at some scale below $M_{Pl}$ -- of a given parameter point depends greatly on the order at which $\lambda(m_t)$ has been extracted from the Higgs mass. Moreover, this is not only a matter of using a loop-corrected matching instead of a tree-level one, but the loop order at which it is performed does also matter.

\subsection{Vector-like quarks and stability of the SM}
\label{sec:VL}
From the SM, it is known that the quartic coupling $\lambda$ runs negative at a scale $Q\simeq 10^9-10^{11}\,$GeV, leading to a metastable but long-lived vacuum \cite{Degrassi:2012ry,Buttazzo:2013uya}. While extensions with a heavy singlet similar to the previous subsections can have a stabilising effect on the potential  \cite{EliasMiro:2012ay,Lebedev:2012zw}, fermionic extensions typically have the opposite effect through the negative impact of the vector-like (VL) fermions Yukawa coupling on the running of $\lambda$, see e.g. \cite{EliasMiro:2011aa,Bambhaniya:2016rbb}. A model where the latter is compensated by the effect of the former is discussed in Ref.~\cite{,Xiao:2014kba}.

\begin{figure}[tb]
\includegraphics[width=\linewidth]{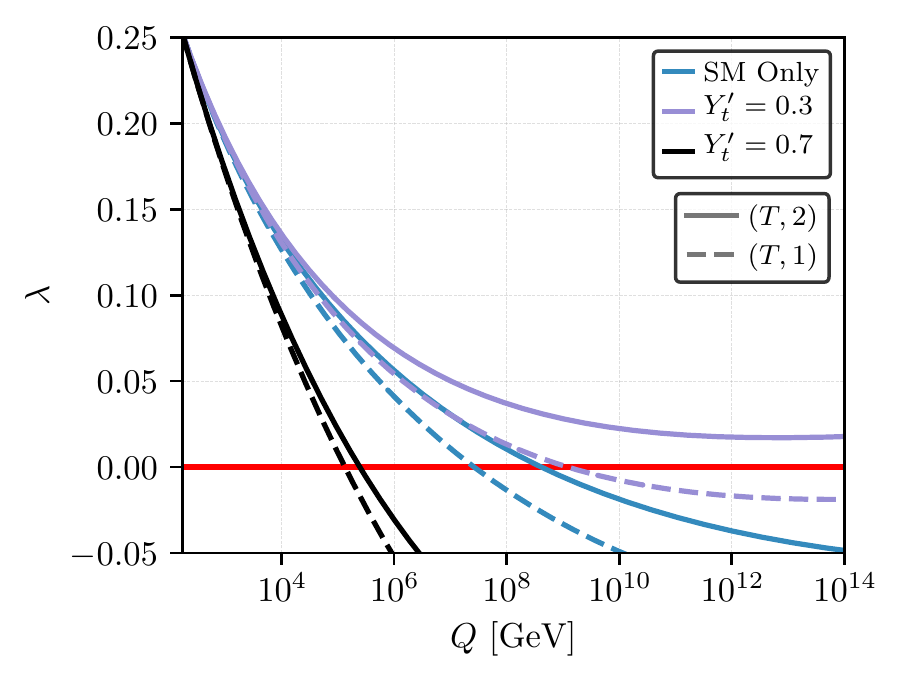}
\caption{Simplified comparison between the running of $\lambda$ in the SM with and without vector-like states. Here, we used full one-loop (dashed lines) and two-loop RGEs (full lines) in both models and as a starting point
the SM best-fit values from Ref.~\cite{Buttazzo:2013uya}. For the purple (black) lines we use $Y_t'=0.3~(0.7)$. 
}
\label{fig:RunVL}
\end{figure}

Here, we shall extend the SM by one generation of a VL quark doublet $Q'$ as well as an 
up-type quark singlet $t'$ with their corresponding counterparts $\tilde Q',\,\tilde t'$, with quantum numbers under the SM gauge group of $t': (\bf{\bar 3},\bf{1},-\frac23)$, $\tilde t': (\bf{3},\bf{1},\frac23)$,   $Q': (\bf{3},\bf{2},\frac16)$, $\tilde Q': (\bf{\bar 3},\bf{2},-\frac16)$. The Lagrangian of the model reads (in terms of two-component spinors)
\begin{align}
\mathscr{L} = \mathscr{L}_{\rm SM} -\big(& Y_{t}' Q' \cdot H t' + \tilde{Y}_{t}' \tilde{Q}' \cdot \overline{H} \tilde{t}' \, \label{eq:VL:pot} \\
& + \, m_T  \tilde{t}' t' \, + \, m_Q \tilde{Q}'  Q' + \text{h.c.}\big)\,. \nn
\end{align}
%
%
For simplicity we take $m_Q = m_T \equiv M_Q$;  we then find at one loop that, with the normalisation of the Higgs quartic coupling $ \mathscr{L} \supset -\frac{1}{2} \lambda |H|^4$
\begin{align}
\lambda_{\rm SM}  &= \lambda_{\rm VLQ}    \label{eq:VL:shift_lambda_analytical} \\
&- \frac{1}{16\pi^2} \bigg[ ( Y_t' + \tilde{Y}_{t}^{\prime})^2 (5 Y_t^{\prime\,2} -2 Y_t' \tilde{Y}_{t}^{\prime} + 5\tilde{Y}_t^{\prime\,2}) \nn\\
& \qquad\qquad + 6 (Y_t^{\prime\,4}+ \tilde{Y}_t^{\prime\,4}) \log \frac{M_Q^2}{\mu^2} \bigg] \nn\\
& + \frac{2 \lambda_{\rm VLQ}}{16\pi^2} \bigg[ (Y_t' - \tilde{Y}_{t}^{\prime})^2 + 3 (Y_t^{\prime\,2} + \tilde{Y}_t^{\prime\,2}) \log \frac{M_Q^2}{\mu^2} \bigg]\,. \nn
\end{align}
Let us first consider the impact of the new vector-like states on the running quartic Higgs coupling. For simplicity, we consider here and in the following examples only one extra non-zero Yukawa interaction $Y_t'$ and consequently set $\tilde Y_t' = 0$ as it does not play a role in the following discussion.\footnote{Although this leads to a stable lightest VL quark, there could for instance be couplings to a hidden sector, leading to a relaxation of the direct collider constraints.} Then for matching at $\mu = m_t$ with $M_Q < $ TeV, the shifts to $\lambda$ are less than $10 \%$ for $Y_t' \lesssim 0.7$, but grow rapidly to $\sim 50\%$ for $Y_t' \sim 1.$ On the other hand, the direct impact of $Y_t'$ on the running of $\lambda$ at one loop is given by 
\begin{align}
16 \pi^2 \beta_\lambda^{(1)} \supset 12 Y_t^{\prime\,2}(\lambda - Y_t^{\prime\,2}) \,,
\label{eq:VL:Ytprime_impact_on_lambda}
\end{align}
which contributes significantly to the negative slope of $\lambda$ for large values but plays a negligible role when $Y_t'$ is small. In the latter case, the impact of the new fermions on the running of the gauge couplings may outweigh their direct impact on $\lambda$. Consider the potential of Eq.~(\ref{eq:VL:pot}). Due to the additional coloured fermions, the running of $g_3$ changes at one-loop to 
\begin{equation}
16 \pi^2 \beta_{g_3}^{(1)} =  \left(- 7 + \frac43 n_T + \frac23 n_Q \right) g_3^3 \to -5 g_3^3\,,
\label{eq:VL:beta_g3}
\end{equation}
i.e. it decreases more slowly when increasing the scale compared to the SM.
In addition, we also obtain a shift in $\alpha^{\MS}_S(m_t)$ of
\begin{equation}
   \alpha^{\MS}_S \to \frac{\alpha^{\MS}_S}{1-\frac{\alpha^{\MS}_S}{\pi} \log(M_Q/m_t)}
   \label{eq:VL:shift_in_alphaS}
  \end{equation}
with respect to the SM. In total, both effects increase the influence of the strong force on the running of $\lambda$, adding positively to the slope.
The impact on  $\lambda$ is shown in Fig.~\ref{fig:RunVL} where the running $\lambda$ is computed using two-loop RGEs 
when assuming the pure SM (blue) and the VL extension (purple and black). 
No matching was applied yet here (i.e. also the shift in $\alpha_S$ was neglected) -- the changes in the VL case therefore entirely stem from the altered running of the gauge couplings, most importantly Eq.~(\ref{eq:VL:beta_g3}). As starting value for $\lambda$ we used  the best-fit value from Ref.~\cite{Buttazzo:2013uya}.
The increased $g_3$ throughout the energy scales leads to a positive contribution to the slope of $\lambda$.
As is seen, it can even lead to a stabilisation of the potential at high energies as long as the direct impact of $Y_t'$ is kept under control by taking it small. This is seen in the purple curve where we have chosen $Y_t'=0.3$. For larger values, the known destabilising effect can overcome the stabilisation from $g_3$. As a consequence, the scale of metastability would coincide with the SM for values of $Y_t' \sim 0.5$ and decreases quickly with larger values. This is also shown in the figure for $Y_{t}'=0.7$ (black line) where 
 $\lambda$ enters the metastable region already at energies of $\sim 10^5\,$GeV. 
We remark that the inclusion of the shift in $\alpha_S$ according to  Eq.~(\ref{eq:VL:shift_in_alphaS}) would lead to an even milder running of $\lambda$. In fact, just using the one-loop RGEs for the case $Y_t'=0.3$ the quartic coupling stays positive over the entire energy range.
 We will discuss the effects of the proper matching, including the shifts in $\lambda$, in what follows. 
 
\begin{figure}[tb]
\includegraphics[width=\linewidth]{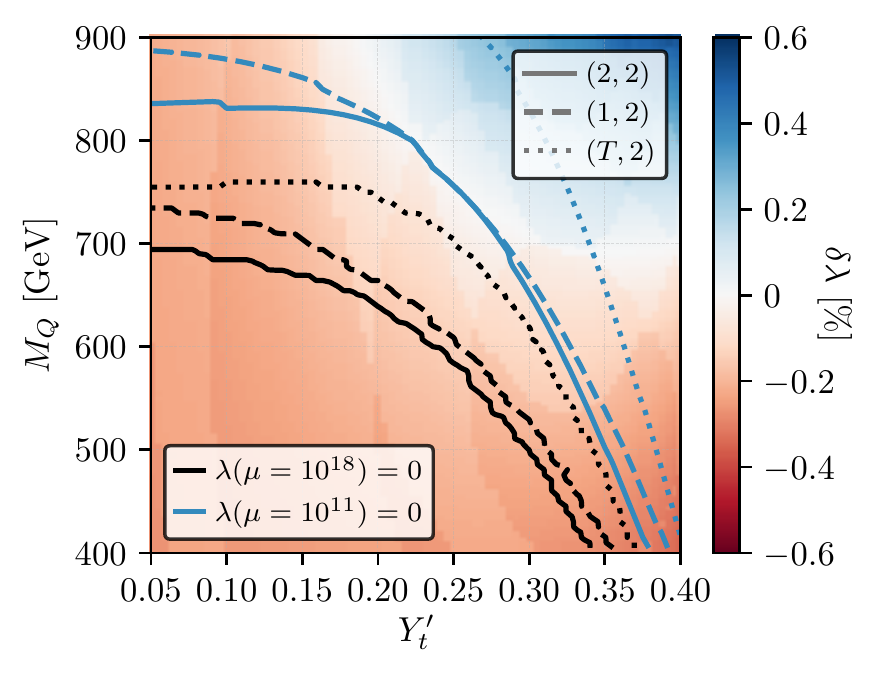}
\caption{Contours of the scale $\Lambda_0$ at which $\lambda$ runs negative with full two-loop RGEs and one-loop (dashed) or two-loop (full lines) matching, and for comparison when using the best-fit value $\lambda^{\rm SM}=0.25208$ at the top mass scale (dotted). Black lines correspond to $\Lambda_0=10^{18}\,$GeV, and blue lines to $10^{11}$\,GeV. The background shows the two-loop shift in $\lambda (m_t)$ in percent, defined as $(\lambda^{(2)} -  \lambda^{(1)})/\lambda^{(1)}$. }
\label{fig:VLloop_metastability_scales}
\end{figure}

Solving Eq.~(\ref{eq:VL:shift_lambda_analytical}) for the matched $\lambda_{\rm VLQ}$ at $\mu=m_t$ and keeping $\tilde Y_t'=0$, we see that the shifts are
slightly negative for small $Y_t' \lesssim 0.45$ and positive for larger Yukawa couplings. This has as a consequence that for low $Y_t'$ where the VL quarks 
help increasing the scale of metastability, loop corrections have the opposite effect. However, the size of the shifts stays below 1\,\%.
In Fig.~\ref{fig:VLloop_metastability_scales}, the contour lines represent the predictions for the scale of metastability as a function of $Y_t'$ and $M_Q= m_Q=m_T$ using tree-level (dotted), one-loop (dashed) as well as two-loop matching (full lines) while applying two-loop RGEs. The colour code in the background quantifies the relative two-loop shift in $\lambda (m_t)$, $(\lambda^{(2)} -  \lambda^{(1)})/\lambda^{(1)}$, which stays below roughly half a percent. Nevertheless, the impact of these small shifts is non-negligible: the corresponding $Y_t'$ values at which $\lambda$ crosses zero at a given scale typically change by more than 10\,\% between tree- and two-loop matching. That is, the scale at which metastability occurs is very sensitive to the starting value of $\lambda$ -- meaning that matching is absolutely crucial to make reliable statements.
After including the correct shifts in $\lambda$, the picture nevertheless remains the same as that for small $Y_t'$; the impact of the VL quarks on $\alpha_S$ 
can be such that the scale of metastability is \emph{increased} with respect to the SM, leading to the possibility of absolute stability all the way up to the Planck scale.

\begin{figure}[tb]
\includegraphics[width=\linewidth]{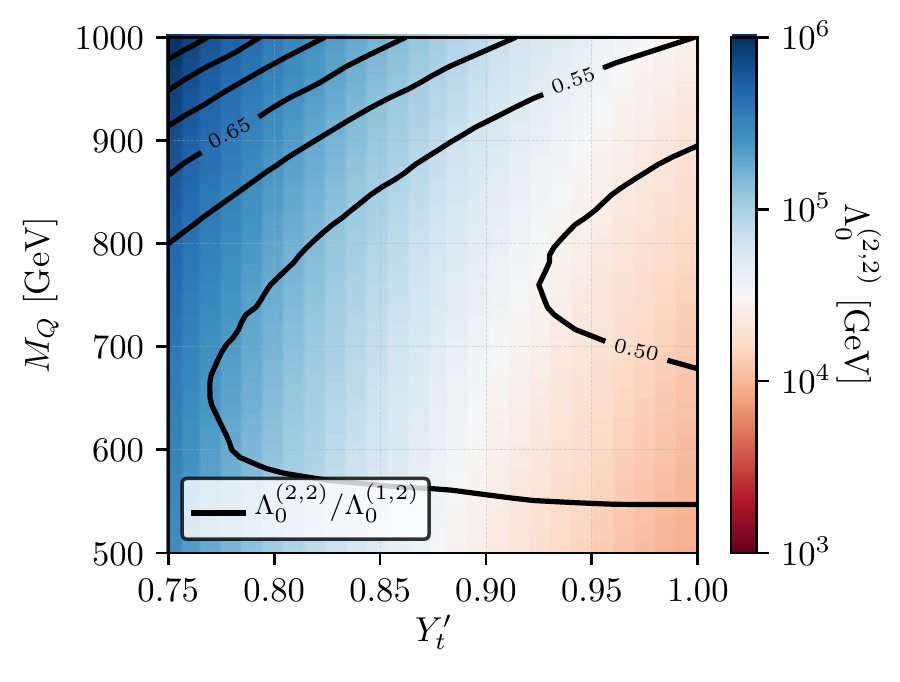}
\caption{The scale of metastability $\Lambda_0$ in the case of large $Y_t'$ using two-loop running with two-loop matching for $\lambda$. The black contours show the size of $\Lambda_0$ with respect to using one-loop matching.}
\label{fig:VLloop2}
\end{figure}

The situation is reversed if $Y_t'$ is large. In that case, we enter the known scenario in which the additional impact of $Y_t'$ on the RGEs
of $\lambda$ drives it negative faster when compared to the SM, further destabilising the vacuum. We show this in Fig.~\ref{fig:VLloop2}. The background shading indicates the scale $\Lambda_0$ at which $\lambda$ crosses zero as a function of $Y_t'$ and $M_Q$, using two-loop matching, whereas the contour lines
represent the relative changes with respect to using one-loop matching, $\Lambda_0^{(2,2)}/\Lambda_0^{(1,2)}$.
As expected, $\Lambda_0$ is well below the pure SM prediction, and becomes smaller for larger $Y_t'$ and smaller $M_Q$.
The differences in $\Lambda_0$ between one- and two-loop matching are quite mild here -- ``only'' $\mathcal O(100\,\%)$ -- since the two-loop corrections to 
$\lambda$ are small. We remark however, that going one order lower and comparing tree-level with one-loop matching using one-loop RGEs, we would see up 
to an order of magnitude differences in the eventual scale $\Lambda_0$.

Summarising, we have shown that vector-like quarks can have both a destabilising but also a stabilising effect on the SM Higgs potential, and that 
the correct inclusion of threshold effects is crucial for obtaining precise predictions about the fate of the electroweak vacuum.

\subsection{Two-Higgs Doublet model}
\label{subsec:THDM}

Finally, as a last example we study the impact of loop-level matching on Two-Higgs Doublet models.\footnote{For an overview, see for instance Ref.~\cite{Branco:2011iw}.} Here we will restrict ourselves to the CP-conserving version with a softly broken $\mathbb Z_2$ symmetry. The corresponding scalar potential can be written as
\begin{align}
V & =  m_1^2\, \Phi_1^\dagger \Phi_1 + m_2^2\, \Phi_2^\dagger \Phi_2 + \lambda_1\, (\Phi_1^\dagger \Phi_1)^2 + \lambda_2\, (\Phi_2^\dagger \Phi_2)^2  \nonumber \\
& \hspace{1cm} + \lambda_3\, (\Phi_1^\dagger \Phi_1) (\Phi_2^\dagger \Phi_2) + \lambda_4\, (\Phi^\dagger_2 \Phi_1)(\Phi^\dagger_1 \Phi_2) \label{eq:scalar_potential} \\ 
& +  M_{12}^2 \,(\Phi_1^\dagger \Phi_2 + \Phi_2^\dagger \Phi_1) + \frac{\lambda_5}{2}\, \left( (\Phi_2^\dagger \Phi_1)^2  + (\Phi_1^\dagger \Phi_2)^2 \right)\,. \nonumber
\end{align} 
Note that our sign convention for $M_{12}^2$ differs from most definitions in the literature.
After electroweak symmetry breaking, we decompose the scalar fields according to
\begin{align}
\Phi_k = \begin{pmatrix}
\phi^+_k \\ 
\frac{1}{\sqrt{2}} (v_k + \phi^0_k + i\,\sigma_k)
\end{pmatrix}\,,~~~i=1,2\,,
\label{eq:THDM:vevs}
\end{align}
where $v_1^2+v_2^2=v^2$ and we define $t_\beta=\tan\beta=v_2/v_1$. 
The charged (neutral CP-odd) fields mix to one physical charged Higgs $H^\pm$ (pseudoscalar $A$) and the corresponding would-be Goldstone bosons. At LO, the angle
$\beta$ coincides with the mixing angle in the pseudoscalar and charged Higgs sector. In the CP-even sector, there are two fields which mix to one light and one heavy eigenstate, with masses $m_h$ and $m_H$.

In the same fashion as for the models used above, we can relate the masses and mixing angles to the quartic couplings, leading to the following relations:
\begin{align}
 \lambda_1 &= \frac{1 + t_\beta^2}{2 (1 + t_\alpha^2) v^2} \left(m_h^2 t_\alpha^2 + m_H^2 + M^2_{12} t_\beta (1+t_\alpha^2) \right) , \label{eq:THDM:couplings_relations_1}\\
 \lambda_2  &= \frac{M_{12}^2(1 + t_\beta^2)}{2 t_\beta^3 v^2} +\frac{(1 + t_\beta^2)\left(m_h^2  + m_H^2 t_\alpha^2 \right)}{2 t_\beta^2 (1 + t_\alpha^2)  v^2}  \,, \label{eq:THDM:couplings_relations_2}\\
 \lambda_3  &= \frac{1}{(1 + t_\alpha^2) t_\beta v^2} \Big[\left(m_H^2-m_h^2\right) t_\alpha (1 + t_\beta^2)  \nonumber \\ &+ 2 m_{H^\pm}^2 (1 + t_\alpha^2) t_\beta + M^2_{12} (1 + t_\alpha^2) (1 + t_\beta^2)\Big]  \,, \label{eq:THDM:couplings_relations_3}\\
 \lambda_4  &=\frac{1}{t_\beta v^2}\left(-M^2_{12}(1+t_\beta^2) + m_A^2 t_\beta - 2 m_{H^\pm}^2 t_\beta\right), \label{eq:THDM:couplings_relations_4} \\
\lambda_5  &= \frac{1}{t_\beta v^2}\left(-M^2_{12}(1+t_\beta^2) - m_A^2 t_\beta \right). \label{eq:THDM:couplings_relations_5}
\end{align}

Analogously to the singlet extension of the SM (Sec~\ref{SEC:singletext}), we define the cut-off scale of a particular scenario as the scale at which either
one of the $\lambda_i$ becomes larger than $4\,\pi$ or the unitarity constraints using the running couplings are violated. The latter are too long to show here but can e.g. easily be computed using the {\tt SARAH} implementation of the model in conjunction with Appendix~D of Ref.~\cite{Krauss:2017xpj}.

\begin{figure}[htbp]
\includegraphics[width=\linewidth]{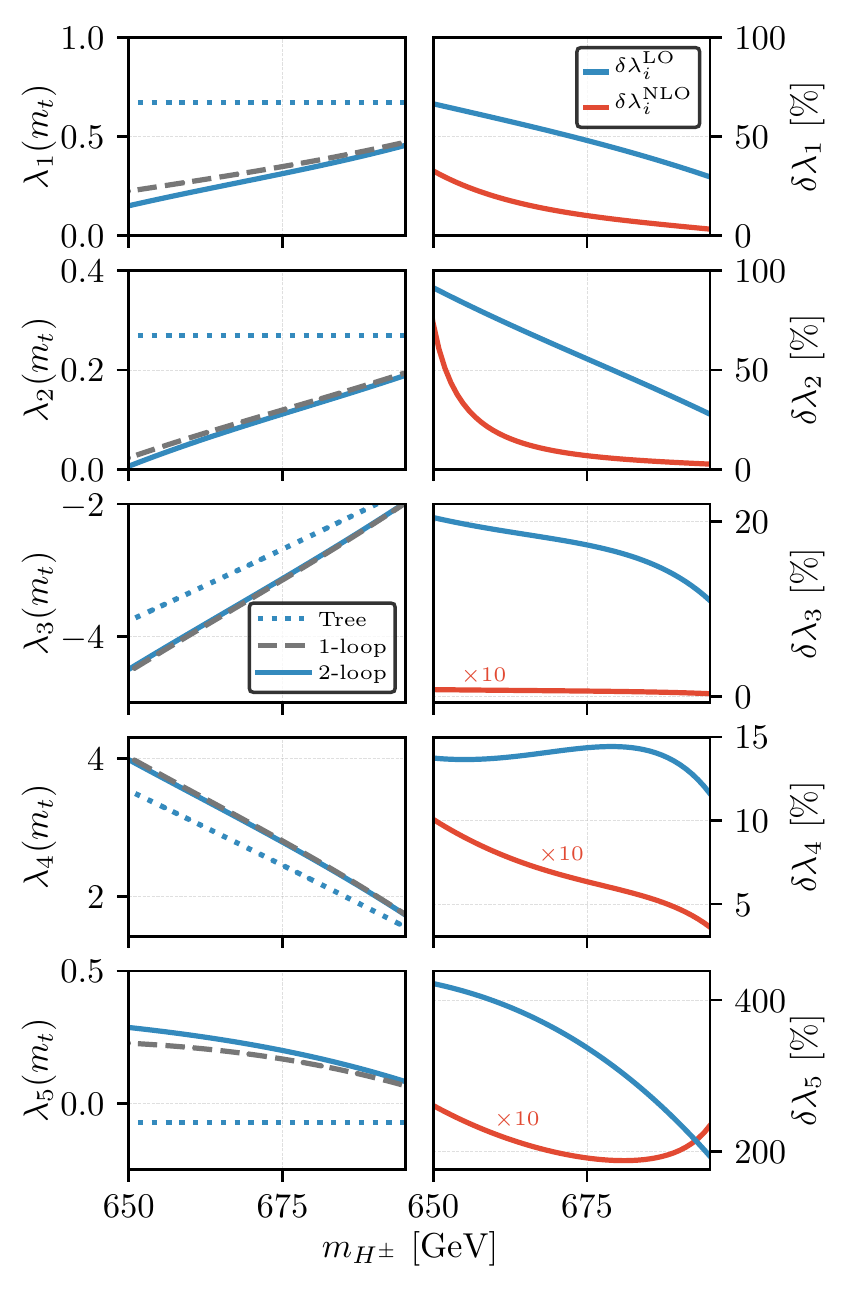}
\caption{The size of the quartic couplings at $m_t$ at LO (blue dotted), NLO (grey dashed) and NNLO (blue solid) as a function of $m_{H^\pm}$. Here, we set the other physical masses and mixing angle to $m_h=125$~GeV, $m_H=750$~GeV, $m_A=730$~GeV and $\tan\alpha=-0.71$. The other model parameters are $\tan\beta$=1.4 and $M_{12}^2=-500^2~\text{GeV}^2$.
On the right, we show the relative differences $\delta \lambda_i^{\rm LO}\equiv |(\lambda_i^{(T)}-\lambda_i^{(1)})/\lambda_i^{(T)}|$ and $\delta \lambda_i^{\rm NLO}\equiv |(\lambda_i^{(1)}-\lambda_i^{(2)})/\lambda_i^{(1)}|$. Note that for $\lambda_i$ where $i=3,4,5$ the NLO differences have been multiplied by an additional factor of 10 to increase visibility.
}
\label{fig:THDMmatch}
\end{figure}

First we are going to look at the matching at the top mass scale. It has already been pointed out in Ref.~\cite{Braathen:2017izn} that the loop corrections to the mass spectrum of THDMs can be significant. In Fig.~\ref{fig:THDMmatch}, we show on the left-hand side the size of the individual couplings $\lambda_i$ for the three matching orders as a function of the charged Higgs mass. 
The leading-order $\lambda_i$ are simple linear functions of this mass according to Eqs.~(\ref{eq:THDM:couplings_relations_1}) to (\ref{eq:THDM:couplings_relations_5}), whereas the $\lambda_i$ evaluated with higher-order matching contain the shifts due to self-energy and tadpole corrections.
We see that large differences of $\mathcal O(100\,\%)$ or even larger can appear between leading and next-to-leading order. The size of the relative shifts is displayed on the right-hand side of each panel. As expected for a converging perturbative series,
the differences between one- and two-loop matching are much less pronounced; however they can still range around tens of percent. Obviously these large differences necessarily have a significant effect on the validity of the theory at higher scales. In the following we will therefore investigate the changes
in cut-off scales between the different approaches.

\begin{figure}[htbp]
\includegraphics[width=\linewidth]{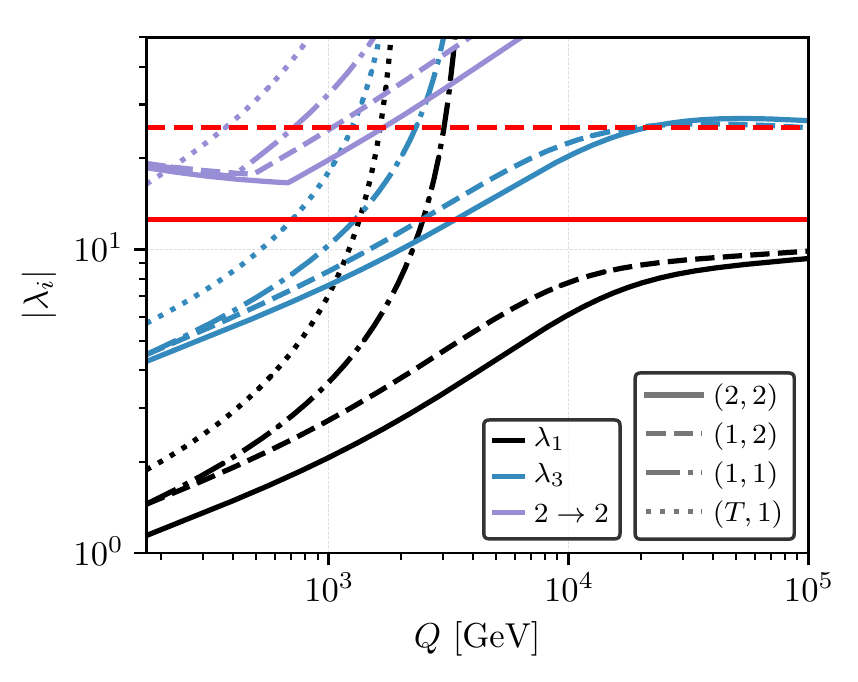}
\caption{RGE running of the individual quartic couplings $\lambda_1$ (black) and $\lambda_3$ (blue) for the parameter point defined by 
$m_H=511\,$GeV, $m_A=607\,$GeV, $m_{H^{\pm}}=605\,$GeV, $t_\beta=1.45$, $t_\alpha=-0.82$, $M_{12}^2=-(250\,{\rm GeV})^2$, 
using $n$-loop level matching and $m$-loop RGEs. The dotted lines stand for $(T,1)$, dot-dashed for (1,1), dashed for (1,2) and full lines for (2,2). Coupling values of of $\pm 4\,\pi$ are indicated by a red solid line. 
In addition, we display the largest eigenvalue of the scalar $2\to 2$ scattering amplitude in purple. The upper bound of $8\pi$ is indicated by a dashed red line. We used the Yukawa scheme of type II.
}
\label{fig:THDM_running_benchmark}
\end{figure}

As mentioned in the Introduction, the two-loop RGEs are well known but often neglected in the literature -- although it is known that large differences can appear; see e.g. Ref.~\cite{Chowdhury:2015yja}. Similar to the singlet-extended SM, the two-loop RGEs tend to moderate the one-loop running. As a result, Landau poles typically appear at much higher scales when including the two-loop effects. For instance, for both $i=1$ and 2,  $16\pi^2 \beta_{\lambda_{i}} \supset 24 \, \lambda_i^2 (1 - \frac{13}{16\pi^2}  \lambda_i) $. The two-loop contribution thereby counteracts the large one-loop slope, stalling the evolution for $\lambda_i$ just below $4\,\pi$.
In contrast, the two-loop RGEs to $\lambda_3$ for instance do have a mitigating effect on the evolution; however a complete stalling only occurs for values much larger than $4\pi$. 
In Fig.~\ref{fig:THDM_running_benchmark} we show for a particular parameter point the running of the couplings $\lambda_1$ (black) and $\lambda_3$ (blue) for $n$-loop matching and $m$-loop RGEs.  The influence of the two-loop RGEs can be best gauged for the cases $(n,m)=(1,1)$ versus $(1,2)$, displayed as dot-dashed and dashed lines, respectively. It is clearly seen that, while both quartic couplings run into a Landau pole close to $Q=3\,$TeV when using one-loop RGEs, the inclusion of the two-loop terms leads to a significant flattening and therefore splitting between the two cases.

\begin{figure}[tb]
\includegraphics{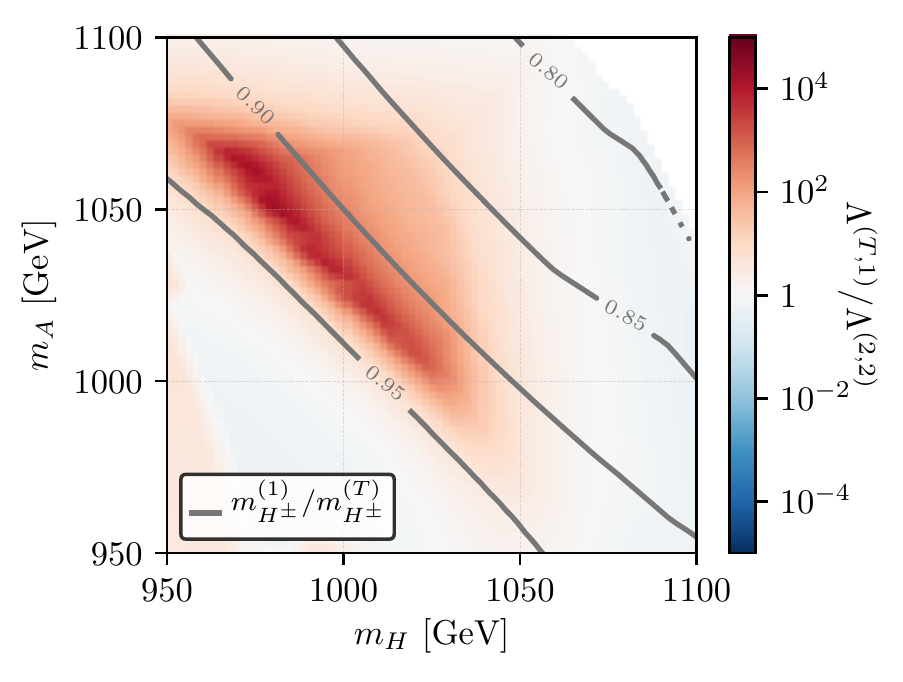}
\caption{Comparison of the cut-off scales between using tree matching and one-loop RGEs and both two-loop matching and RGEs for a THDM region
with large $M_{12}^2=-(750\,{\rm GeV})^2$. The loop-level spectrum was evaluated taking the tree-level values $m_{H^\pm}=1.14\,$TeV and $t_\alpha=-0.95$ as inputs. The ratio of the loop-corrected charged Higgs mass to its tree-level input is shown as grey contour lines. The quartic couplings for the case of tree matching were obtained using the leading-order relations Eqs.~(\ref{eq:THDM:couplings_relations_1}) to (\ref{eq:THDM:couplings_relations_5}), taking the spectrum of the 
two-loop calculation as input. We further fixed $\tan\beta=1.14$ and applied the Yukawa scheme of type I.}
\label{fig:THDM_2Dplane}
\end{figure}

Let us look at the impact of the threshold corrections next. As shown in the example of Fig.~\ref{fig:THDMmatch}, the threshold corrections for the $\lambda_i$  can be significant. This can also be seen in the starting values of the couplings at $m_t$ in Fig.~\ref{fig:THDM_running_benchmark}: the values for $\lambda_1$ using tree-level (one-loop) [two-loop] matching are 1.88 (1.45) [1.14] whereas for $\lambda_3$, the values are 5.7 (4.5) [4.3]. The decrease in value at higher loop orders comes from 
the fact that in this particular scenario, the average loop corrections to the scalar masses are positive. As a result, one obtains a negative shift in the $\lambda_i$ at the matching scale when demanding an on-shell renormalisation scheme. Consequently, the cut-off scale increases with every additional loop order.
Finally, the purple lines show the maximal eigenvalue of the scalar $2\to 2$ scattering matrix as well as the corresponding upper bound of $8\,\pi$ from perturbative unitarity. We see that, while the parameter point would seem to violate perturbative unitarity already at scales below 400\,GeV when using the conventional approach, i.e. tree-level matching with one-loop RGEs, it actually only does so just below $Q=2\,$TeV. Existing studies would have discarded such a parameter point, due to the breakdown of unitarity at energies probeable by the LHC. To that end including matching can result in an increase in the experimental viability of large regions of parameter space.

However, it need not be the case that the cut-off scale is raised by higher loop effects. Indeed, for large values of $|M_{12}|$ and therefore large heavy scalar masses, the mass corrections can be large and negative -- leading to the opposite effects, i.e. a decreased cut-off scale due to larger quartic couplings after the inclusion of the proper matching. An example where this happens is 
presented in Fig.~\ref{fig:THDM_2Dplane}. 
For this figure we have evaluated the spectrum at the two-loop level while fixing the tree-level input values of $t_\alpha$ as well as $m_{H^\pm}$ which enter the spectrum calculation at the loop level. Therefore, the loop-corrected $m_{H^\pm}$ varies over this plane. The grey contours show the ratio of the loop-corrected charged Higgs mass over the tree-level input, $m_{H^\pm}^{(1)}/m_{H^\pm}^{(T)}$. 
To obtain the LO couplings, i.e. the case of tree-level matching, we take the loop-corrected spectrum and calculate $\lambda_i$ according to Eqs.~(\ref{eq:THDM:couplings_relations_1}) to (\ref{eq:THDM:couplings_relations_5}). 
Finally, we run the couplings up in scale using two-loop RGEs for the two-loop- and one-loop RGEs for the tree-level-matched couplings in order to evaluate the cut-off scale.
The coloured contours show the ratio of the cut-off scales, $\Lambda^{(T,1)}/\Lambda^{(2,2)}$, obtained with tree-level matching and one-loop RGEs and with two-loop matching and two-loop RGEs, respectively. 
In particular in the region where all heavy scalar masses are approximately equal, we observe large differences in cut-off scales. In fact, while the 
tree-level matching approach suggests a cut-off at $\mathcal O(10^7\,{\rm GeV})$, the full two-loop matching procedure demands new physics restoring unitarity and perturbativity already at the TeV scale.

Concluding, the conventional approach of tree-level matching and one-loop RGEs can both over- but also underestimate the cut-off scale by many orders of magnitude. It is therefore of crucial importance to (i) take into account the -- known -- RGEs beyond one-loop and to (ii) consistently match the 
couplings before running.



\section{Summary and conclusions}
\label{sec:conclusions}

In this paper, we have investigated the impact that matching plays in the high-scale validity of minimal extensions of the Standard Model (SM). We argued that the usual approach of using $N-1$ matching when utilising $N$-loop RGEs neglects important contributions in the presence of large couplings. 
In fact, for most non-supersymmetric models, studies beyond tree-level matching and one-loop RGEs are rare or even absent.
We analysed in different scenarios the impact of both matching at two-loop order as well
as the two-loop RGEs, highlighting the differences with respect to previous approaches. For simple models, we provided 
an analytical computation of the matching conditions.
We pointed out how sensitive the cut-off scale of the real-singlet-extended SM is to the loop order of both matching and 
RGE running and showed that the scale dependence decreases for $N$- with respect to $N-1$-loop matching. 
Imposing an additional $\mathbb Z_2$ symmetry on this model furthermore enabled us to study the fate of the electroweak 
vacuum as well as the UV completion analytically. We highlighted regions of parameter space where the model can in principle be 
valid up to the Planck scale -- a statement which crucially depends on the proper matching of the quartic couplings at the low scale.

In a scenario where the SM is extended by vector-like quarks, we showed that the impact of the latter can actually increase the 
Higgs quartic interaction such that it does not become negative at higher scales -- an observation which we have not encountered before in the literature. The reason is that, despite the negative impact of the additional Yukawa coupling 
on the running of $\lambda$, the presence of additional coloured states modifies the running strong coupling in such a way that 
it adds positively to the $\beta$-function of $\lambda$. Also in this scenario, the matching of $\lambda$ before the RGE evolution
has a significant impact on the predicted high-scale behaviour of the model.

As a final example we showed in a Two-Higgs Doublet model that the loop-level matching of the quartic couplings can lead to significant changes in both the \MS values and subsequently the cut-off scale of the theory.

To conclude, we observe that robust statements about the UV behaviour of non-supersymmetric, weakly coupled BSM models can only be made when including, at the very least, loop-level matching. We stress that the required loop-level corrections, as well as two-loop RGEs, are readily accessible with the computer tool {\tt SARAH} for any general renormalisable field theory. In light of our results, we strongly encourage its use when accurate high-scale predictions are required.

\section*{Acknowledgements}
MEK, TO and FS thank the LPTHE in Paris for their hospitality while part of this work was completed. JB and MDG acknowledge support from French state funds managed by the Agence Nationale de la Recherche (ANR), in the context of the LABEX ILP (ANR-11-IDEX-0004-02, ANR-10-LABX-63), and MDG acknowledges support from the ANR grant ``HiggsAutomator'' (ANR-15-CE31-0002). JB was supported by a scholarship from the Fondation CFM. MEK is supported by the DFG Research Unit 2239 ``New Physics at the LHC''. The work of TO has been supported by the German Research Foundation (DFG) under grant Nos. EXC-1098, FOR 2239 and SFB/Transregio 33 ``The Dark Universe'' as well as by the European Research Council (ERC) under the European Union's Horizon 2020 research and innovation programme (grant agreement No. 637506, ``$\nu$Directions'').
FS is supported by the ERC Recognition Award ERC-RA-0008 of the Helmholtz Association. 

\bibliography{literature}

\begin{thebibliography}{49}%
\makeatletter
\providecommand \@ifxundefined [1]{%
 \@ifx{#1\undefined}
}%
\providecommand \@ifnum [1]{%
 \ifnum #1\expandafter \@firstoftwo
 \else \expandafter \@secondoftwo
 \fi
}%
\providecommand \@ifx [1]{%
 \ifx #1\expandafter \@firstoftwo
 \else \expandafter \@secondoftwo
 \fi
}%
\providecommand \natexlab [1]{#1}%
\providecommand \enquote  [1]{``#1''}%
\providecommand \bibnamefont  [1]{#1}%
\providecommand \bibfnamefont [1]{#1}%
\providecommand \citenamefont [1]{#1}%
\providecommand \href@noop [0]{\@secondoftwo}%
\providecommand \href [0]{\begingroup \@sanitize@url \@href}%
\providecommand \@href[1]{\@@startlink{#1}\@@href}%
\providecommand \@@href[1]{\endgroup#1\@@endlink}%
\providecommand \@sanitize@url [0]{\catcode `\\12\catcode `\$12\catcode
  `\&12\catcode `\#12\catcode `\^12\catcode `\_12\catcode `\%12\relax}%
\providecommand \@@startlink[1]{}%
\providecommand \@@endlink[0]{}%
\providecommand \url  [0]{\begingroup\@sanitize@url \@url }%
\providecommand \@url [1]{\endgroup\@href {#1}{\urlprefix }}%
\providecommand \urlprefix  [0]{URL }%
\providecommand \Eprint [0]{\href }%
\providecommand \doibase [0]{http://dx.doi.org/}%
\providecommand \selectlanguage [0]{\@gobble}%
\providecommand \bibinfo  [0]{\@secondoftwo}%
\providecommand \bibfield  [0]{\@secondoftwo}%
\providecommand \translation [1]{[#1]}%
\providecommand \BibitemOpen [0]{}%
\providecommand \bibitemStop [0]{}%
\providecommand \bibitemNoStop [0]{.\EOS\space}%
\providecommand \EOS [0]{\spacefactor3000\relax}%
\providecommand \BibitemShut  [1]{\csname bibitem#1\endcsname}%
\let\auto@bib@innerbib\@empty
\bibitem [{\citenamefont {Buttazzo}\ \emph {et~al.}(2013)\citenamefont
  {Buttazzo}, \citenamefont {Degrassi}, \citenamefont {Giardino}, \citenamefont
  {Giudice}, \citenamefont {Sala}, \citenamefont {Salvio},\ and\ \citenamefont
  {Strumia}}]{Buttazzo:2013uya}%
  \BibitemOpen
  \bibfield  {author} {\bibinfo {author} {\bibfnamefont {D.}~\bibnamefont
  {Buttazzo}}, \bibinfo {author} {\bibfnamefont {G.}~\bibnamefont {Degrassi}},
  \bibinfo {author} {\bibfnamefont {P.~P.}\ \bibnamefont {Giardino}}, \bibinfo
  {author} {\bibfnamefont {G.~F.}\ \bibnamefont {Giudice}}, \bibinfo {author}
  {\bibfnamefont {F.}~\bibnamefont {Sala}}, \bibinfo {author} {\bibfnamefont
  {A.}~\bibnamefont {Salvio}}, \ and\ \bibinfo {author} {\bibfnamefont
  {A.}~\bibnamefont {Strumia}},\ }\href {\doibase 10.1007/JHEP12(2013)089}
  {\bibfield  {journal} {\bibinfo  {journal} {JHEP}\ }\textbf {\bibinfo
  {volume} {12}},\ \bibinfo {pages} {089} (\bibinfo {year} {2013})},\ \Eprint
  {http://arxiv.org/abs/1307.3536} {arXiv:1307.3536 [hep-ph]} \BibitemShut
  {NoStop}%
\bibitem [{\citenamefont {Degrassi}\ \emph {et~al.}(2012)\citenamefont
  {Degrassi}, \citenamefont {Di~Vita}, \citenamefont {Elias-Miro},
  \citenamefont {Espinosa}, \citenamefont {Giudice}, \citenamefont {Isidori},\
  and\ \citenamefont {Strumia}}]{Degrassi:2012ry}%
  \BibitemOpen
  \bibfield  {author} {\bibinfo {author} {\bibfnamefont {G.}~\bibnamefont
  {Degrassi}}, \bibinfo {author} {\bibfnamefont {S.}~\bibnamefont {Di~Vita}},
  \bibinfo {author} {\bibfnamefont {J.}~\bibnamefont {Elias-Miro}}, \bibinfo
  {author} {\bibfnamefont {J.~R.}\ \bibnamefont {Espinosa}}, \bibinfo {author}
  {\bibfnamefont {G.~F.}\ \bibnamefont {Giudice}}, \bibinfo {author}
  {\bibfnamefont {G.}~\bibnamefont {Isidori}}, \ and\ \bibinfo {author}
  {\bibfnamefont {A.}~\bibnamefont {Strumia}},\ }\href {\doibase
  10.1007/JHEP08(2012)098} {\bibfield  {journal} {\bibinfo  {journal} {JHEP}\
  }\textbf {\bibinfo {volume} {08}},\ \bibinfo {pages} {098} (\bibinfo {year}
  {2012})},\ \Eprint {http://arxiv.org/abs/1205.6497} {arXiv:1205.6497
  [hep-ph]} \BibitemShut {NoStop}%
\bibitem [{\citenamefont {Andreassen}\ \emph {et~al.}(2017)\citenamefont
  {Andreassen}, \citenamefont {Frost},\ and\ \citenamefont
  {Schwartz}}]{Andreassen:2017rzq}%
  \BibitemOpen
  \bibfield  {author} {\bibinfo {author} {\bibfnamefont {A.}~\bibnamefont
  {Andreassen}}, \bibinfo {author} {\bibfnamefont {W.}~\bibnamefont {Frost}}, \
  and\ \bibinfo {author} {\bibfnamefont {M.~D.}\ \bibnamefont {Schwartz}},\
  }\href@noop {} {\  (\bibinfo {year} {2017})},\ \Eprint
  {http://arxiv.org/abs/1707.08124} {arXiv:1707.08124 [hep-ph]} \BibitemShut
  {NoStop}%
\bibitem [{\citenamefont {Spencer-Smith}(2014)}]{Spencer-Smith:2014woa}%
  \BibitemOpen
  \bibfield  {author} {\bibinfo {author} {\bibfnamefont {A.}~\bibnamefont
  {Spencer-Smith}},\ }\href@noop {} {\  (\bibinfo {year} {2014})},\ \Eprint
  {http://arxiv.org/abs/1405.1975} {arXiv:1405.1975 [hep-ph]} \BibitemShut
  {NoStop}%
\bibitem [{\citenamefont {Lerner}\ and\ \citenamefont
  {McDonald}(2009)}]{Lerner:2009xg}%
  \BibitemOpen
  \bibfield  {author} {\bibinfo {author} {\bibfnamefont {R.~N.}\ \bibnamefont
  {Lerner}}\ and\ \bibinfo {author} {\bibfnamefont {J.}~\bibnamefont
  {McDonald}},\ }\href {\doibase 10.1103/PhysRevD.80.123507} {\bibfield
  {journal} {\bibinfo  {journal} {Phys. Rev.}\ }\textbf {\bibinfo {volume}
  {D80}},\ \bibinfo {pages} {123507} (\bibinfo {year} {2009})},\ \Eprint
  {http://arxiv.org/abs/0909.0520} {arXiv:0909.0520 [hep-ph]} \BibitemShut
  {NoStop}%
\bibitem [{\citenamefont {Lebedev}(2012)}]{Lebedev:2012zw}%
  \BibitemOpen
  \bibfield  {author} {\bibinfo {author} {\bibfnamefont {O.}~\bibnamefont
  {Lebedev}},\ }\href {\doibase 10.1140/epjc/s10052-012-2058-2} {\bibfield
  {journal} {\bibinfo  {journal} {Eur. Phys. J.}\ }\textbf {\bibinfo {volume}
  {C72}},\ \bibinfo {pages} {2058} (\bibinfo {year} {2012})},\ \Eprint
  {http://arxiv.org/abs/1203.0156} {arXiv:1203.0156 [hep-ph]} \BibitemShut
  {NoStop}%
\bibitem [{\citenamefont {Elias-Miro}\ \emph
  {et~al.}(2012{\natexlab{a}})\citenamefont {Elias-Miro}, \citenamefont
  {Espinosa}, \citenamefont {Giudice}, \citenamefont {Lee},\ and\ \citenamefont
  {Strumia}}]{EliasMiro:2012ay}%
  \BibitemOpen
  \bibfield  {author} {\bibinfo {author} {\bibfnamefont {J.}~\bibnamefont
  {Elias-Miro}}, \bibinfo {author} {\bibfnamefont {J.~R.}\ \bibnamefont
  {Espinosa}}, \bibinfo {author} {\bibfnamefont {G.~F.}\ \bibnamefont
  {Giudice}}, \bibinfo {author} {\bibfnamefont {H.~M.}\ \bibnamefont {Lee}}, \
  and\ \bibinfo {author} {\bibfnamefont {A.}~\bibnamefont {Strumia}},\ }\href
  {\doibase 10.1007/JHEP06(2012)031} {\bibfield  {journal} {\bibinfo  {journal}
  {JHEP}\ }\textbf {\bibinfo {volume} {06}},\ \bibinfo {pages} {031} (\bibinfo
  {year} {2012}{\natexlab{a}})},\ \Eprint {http://arxiv.org/abs/1203.0237}
  {arXiv:1203.0237 [hep-ph]} \BibitemShut {NoStop}%
\bibitem [{\citenamefont {Pruna}\ and\ \citenamefont
  {Robens}(2013)}]{Pruna:2013bma}%
  \BibitemOpen
  \bibfield  {author} {\bibinfo {author} {\bibfnamefont {G.~M.}\ \bibnamefont
  {Pruna}}\ and\ \bibinfo {author} {\bibfnamefont {T.}~\bibnamefont {Robens}},\
  }\href {\doibase 10.1103/PhysRevD.88.115012} {\bibfield  {journal} {\bibinfo
  {journal} {Phys. Rev.}\ }\textbf {\bibinfo {volume} {D88}},\ \bibinfo {pages}
  {115012} (\bibinfo {year} {2013})},\ \Eprint {http://arxiv.org/abs/1303.1150}
  {arXiv:1303.1150 [hep-ph]} \BibitemShut {NoStop}%
\bibitem [{\citenamefont {Costa}\ \emph {et~al.}(2015)\citenamefont {Costa},
  \citenamefont {Morais}, \citenamefont {Sampaio},\ and\ \citenamefont
  {Santos}}]{Costa:2014qga}%
  \BibitemOpen
  \bibfield  {author} {\bibinfo {author} {\bibfnamefont {R.}~\bibnamefont
  {Costa}}, \bibinfo {author} {\bibfnamefont {A.~P.}\ \bibnamefont {Morais}},
  \bibinfo {author} {\bibfnamefont {M.~O.~P.}\ \bibnamefont {Sampaio}}, \ and\
  \bibinfo {author} {\bibfnamefont {R.}~\bibnamefont {Santos}},\ }\href
  {\doibase 10.1103/PhysRevD.92.025024} {\bibfield  {journal} {\bibinfo
  {journal} {Phys. Rev.}\ }\textbf {\bibinfo {volume} {D92}},\ \bibinfo {pages}
  {025024} (\bibinfo {year} {2015})},\ \Eprint {http://arxiv.org/abs/1411.4048}
  {arXiv:1411.4048 [hep-ph]} \BibitemShut {NoStop}%
\bibitem [{\citenamefont {Hamada}\ \emph {et~al.}(2015)\citenamefont {Hamada},
  \citenamefont {Kawana},\ and\ \citenamefont {Tsumura}}]{Hamada:2015bra}%
  \BibitemOpen
  \bibfield  {author} {\bibinfo {author} {\bibfnamefont {Y.}~\bibnamefont
  {Hamada}}, \bibinfo {author} {\bibfnamefont {K.}~\bibnamefont {Kawana}}, \
  and\ \bibinfo {author} {\bibfnamefont {K.}~\bibnamefont {Tsumura}},\ }\href
  {\doibase 10.1016/j.physletb.2015.05.072} {\bibfield  {journal} {\bibinfo
  {journal} {Phys. Lett.}\ }\textbf {\bibinfo {volume} {B747}},\ \bibinfo
  {pages} {238} (\bibinfo {year} {2015})},\ \Eprint
  {http://arxiv.org/abs/1505.01721} {arXiv:1505.01721 [hep-ph]} \BibitemShut
  {NoStop}%
\bibitem [{\citenamefont {Khan}(2016)}]{Khan:2016sxm}%
  \BibitemOpen
  \bibfield  {author} {\bibinfo {author} {\bibfnamefont {N.}~\bibnamefont
  {Khan}},\ }\href@noop {} {\  (\bibinfo {year} {2016})},\ \Eprint
  {http://arxiv.org/abs/1610.03178} {arXiv:1610.03178 [hep-ph]} \BibitemShut
  {NoStop}%
\bibitem [{\citenamefont {Cheon}\ and\ \citenamefont
  {Kang}(2013)}]{Cheon:2012rh}%
  \BibitemOpen
  \bibfield  {author} {\bibinfo {author} {\bibfnamefont {H.~S.}\ \bibnamefont
  {Cheon}}\ and\ \bibinfo {author} {\bibfnamefont {S.~K.}\ \bibnamefont
  {Kang}},\ }\href {\doibase 10.1007/JHEP09(2013)085} {\bibfield  {journal}
  {\bibinfo  {journal} {JHEP}\ }\textbf {\bibinfo {volume} {09}},\ \bibinfo
  {pages} {085} (\bibinfo {year} {2013})},\ \Eprint
  {http://arxiv.org/abs/1207.1083} {arXiv:1207.1083 [hep-ph]} \BibitemShut
  {NoStop}%
\bibitem [{\citenamefont {Chakrabarty}\ \emph {et~al.}(2014)\citenamefont
  {Chakrabarty}, \citenamefont {Dey},\ and\ \citenamefont
  {Mukhopadhyaya}}]{Chakrabarty:2014aya}%
  \BibitemOpen
  \bibfield  {author} {\bibinfo {author} {\bibfnamefont {N.}~\bibnamefont
  {Chakrabarty}}, \bibinfo {author} {\bibfnamefont {U.~K.}\ \bibnamefont
  {Dey}}, \ and\ \bibinfo {author} {\bibfnamefont {B.}~\bibnamefont
  {Mukhopadhyaya}},\ }\href {\doibase 10.1007/JHEP12(2014)166} {\bibfield
  {journal} {\bibinfo  {journal} {JHEP}\ }\textbf {\bibinfo {volume} {12}},\
  \bibinfo {pages} {166} (\bibinfo {year} {2014})},\ \Eprint
  {http://arxiv.org/abs/1407.2145} {arXiv:1407.2145 [hep-ph]} \BibitemShut
  {NoStop}%
\bibitem [{\citenamefont {Chakrabarty}\ and\ \citenamefont
  {Mukhopadhyaya}(2017{\natexlab{a}})}]{Chakrabarty:2016smc}%
  \BibitemOpen
  \bibfield  {author} {\bibinfo {author} {\bibfnamefont {N.}~\bibnamefont
  {Chakrabarty}}\ and\ \bibinfo {author} {\bibfnamefont {B.}~\bibnamefont
  {Mukhopadhyaya}},\ }\href {\doibase 10.1140/epjc/s10052-017-4705-0}
  {\bibfield  {journal} {\bibinfo  {journal} {Eur. Phys. J.}\ }\textbf
  {\bibinfo {volume} {C77}},\ \bibinfo {pages} {153} (\bibinfo {year}
  {2017}{\natexlab{a}})},\ \Eprint {http://arxiv.org/abs/1603.05883}
  {arXiv:1603.05883 [hep-ph]} \BibitemShut {NoStop}%
\bibitem [{\citenamefont {Ferreira}\ \emph {et~al.}(2015)\citenamefont
  {Ferreira}, \citenamefont {Haber},\ and\ \citenamefont
  {Santos}}]{Ferreira:2015rha}%
  \BibitemOpen
  \bibfield  {author} {\bibinfo {author} {\bibfnamefont {P.}~\bibnamefont
  {Ferreira}}, \bibinfo {author} {\bibfnamefont {H.~E.}\ \bibnamefont {Haber}},
  \ and\ \bibinfo {author} {\bibfnamefont {E.}~\bibnamefont {Santos}},\ }\href
  {\doibase 10.1103/PhysRevD.92.033003, 10.1103/PhysRevD.94.059903} {\bibfield
  {journal} {\bibinfo  {journal} {Phys. Rev.}\ }\textbf {\bibinfo {volume}
  {D92}},\ \bibinfo {pages} {033003} (\bibinfo {year} {2015})},\ \bibinfo
  {note} {[Erratum: Phys. Rev.D94,no.5,059903(2016)]},\ \Eprint
  {http://arxiv.org/abs/1505.04001} {arXiv:1505.04001 [hep-ph]} \BibitemShut
  {NoStop}%
\bibitem [{\citenamefont {Chakrabarty}\ and\ \citenamefont
  {Mukhopadhyaya}(2017{\natexlab{b}})}]{Chakrabarty:2017qkh}%
  \BibitemOpen
  \bibfield  {author} {\bibinfo {author} {\bibfnamefont {N.}~\bibnamefont
  {Chakrabarty}}\ and\ \bibinfo {author} {\bibfnamefont {B.}~\bibnamefont
  {Mukhopadhyaya}},\ }\href {\doibase 10.1103/PhysRevD.96.035028} {\bibfield
  {journal} {\bibinfo  {journal} {Phys. Rev.}\ }\textbf {\bibinfo {volume}
  {D96}},\ \bibinfo {pages} {035028} (\bibinfo {year} {2017}{\natexlab{b}})},\
  \Eprint {http://arxiv.org/abs/1702.08268} {arXiv:1702.08268 [hep-ph]}
  \BibitemShut {NoStop}%
\bibitem [{\citenamefont {Chowdhury}\ and\ \citenamefont
  {Eberhardt}(2015)}]{Chowdhury:2015yja}%
  \BibitemOpen
  \bibfield  {author} {\bibinfo {author} {\bibfnamefont {D.}~\bibnamefont
  {Chowdhury}}\ and\ \bibinfo {author} {\bibfnamefont {O.}~\bibnamefont
  {Eberhardt}},\ }\href {\doibase 10.1007/JHEP11(2015)052} {\bibfield
  {journal} {\bibinfo  {journal} {JHEP}\ }\textbf {\bibinfo {volume} {11}},\
  \bibinfo {pages} {052} (\bibinfo {year} {2015})},\ \Eprint
  {http://arxiv.org/abs/1503.08216} {arXiv:1503.08216 [hep-ph]} \BibitemShut
  {NoStop}%
\bibitem [{\citenamefont {Gori}\ \emph {et~al.}(2017)\citenamefont {Gori},
  \citenamefont {Haber},\ and\ \citenamefont {Santos}}]{Gori:2017qwg}%
  \BibitemOpen
  \bibfield  {author} {\bibinfo {author} {\bibfnamefont {S.}~\bibnamefont
  {Gori}}, \bibinfo {author} {\bibfnamefont {H.~E.}\ \bibnamefont {Haber}}, \
  and\ \bibinfo {author} {\bibfnamefont {E.}~\bibnamefont {Santos}},\ }\href
  {\doibase 10.1007/JHEP06(2017)110} {\bibfield  {journal} {\bibinfo  {journal}
  {JHEP}\ }\textbf {\bibinfo {volume} {06}},\ \bibinfo {pages} {110} (\bibinfo
  {year} {2017})},\ \Eprint {http://arxiv.org/abs/1703.05873} {arXiv:1703.05873
  [hep-ph]} \BibitemShut {NoStop}%
\bibitem [{\citenamefont {Basler}\ \emph
  {et~al.}(2017{\natexlab{a}})\citenamefont {Basler}, \citenamefont {Ferreira},
  \citenamefont {Muhlleitner},\ and\ \citenamefont {Santos}}]{Basler:2017nzu}%
  \BibitemOpen
  \bibfield  {author} {\bibinfo {author} {\bibfnamefont {P.}~\bibnamefont
  {Basler}}, \bibinfo {author} {\bibfnamefont {P.~M.}\ \bibnamefont
  {Ferreira}}, \bibinfo {author} {\bibfnamefont {M.}~\bibnamefont
  {Muhlleitner}}, \ and\ \bibinfo {author} {\bibfnamefont {R.}~\bibnamefont
  {Santos}},\ }\href@noop {} {\  (\bibinfo {year} {2017}{\natexlab{a}})},\
  \Eprint {http://arxiv.org/abs/1710.10410} {arXiv:1710.10410 [hep-ph]}
  \BibitemShut {NoStop}%
\bibitem [{\citenamefont {Bhupal~Dev}\ and\ \citenamefont
  {Pilaftsis}(2014)}]{Dev:2014yca}%
  \BibitemOpen
  \bibfield  {author} {\bibinfo {author} {\bibfnamefont {P.~S.}\ \bibnamefont
  {Bhupal~Dev}}\ and\ \bibinfo {author} {\bibfnamefont {A.}~\bibnamefont
  {Pilaftsis}},\ }\href {\doibase 10.1007/JHEP11(2015)147,
  10.1007/JHEP12(2014)024} {\bibfield  {journal} {\bibinfo  {journal} {JHEP}\
  }\textbf {\bibinfo {volume} {12}},\ \bibinfo {pages} {024} (\bibinfo {year}
  {2014})},\ \bibinfo {note} {[Erratum: JHEP11,147(2015)]},\ \Eprint
  {http://arxiv.org/abs/1408.3405} {arXiv:1408.3405 [hep-ph]} \BibitemShut
  {NoStop}%
\bibitem [{\citenamefont {Das}\ and\ \citenamefont {Saha}(2015)}]{Das:2015mwa}%
  \BibitemOpen
  \bibfield  {author} {\bibinfo {author} {\bibfnamefont {D.}~\bibnamefont
  {Das}}\ and\ \bibinfo {author} {\bibfnamefont {I.}~\bibnamefont {Saha}},\
  }\href {\doibase 10.1103/PhysRevD.91.095024} {\bibfield  {journal} {\bibinfo
  {journal} {Phys. Rev.}\ }\textbf {\bibinfo {volume} {D91}},\ \bibinfo {pages}
  {095024} (\bibinfo {year} {2015})},\ \Eprint
  {http://arxiv.org/abs/1503.02135} {arXiv:1503.02135 [hep-ph]} \BibitemShut
  {NoStop}%
\bibitem [{\citenamefont {Xiao}\ and\ \citenamefont {Yu}(2014)}]{Xiao:2014kba}%
  \BibitemOpen
  \bibfield  {author} {\bibinfo {author} {\bibfnamefont {M.-L.}\ \bibnamefont
  {Xiao}}\ and\ \bibinfo {author} {\bibfnamefont {J.-H.}\ \bibnamefont {Yu}},\
  }\href {\doibase 10.1103/PhysRevD.90.014007, 10.1103/PhysRevD.90.019901}
  {\bibfield  {journal} {\bibinfo  {journal} {Phys. Rev.}\ }\textbf {\bibinfo
  {volume} {D90}},\ \bibinfo {pages} {014007} (\bibinfo {year} {2014})},\
  \bibinfo {note} {[Addendum: Phys. Rev.D90,no.1,019901(2014)]},\ \Eprint
  {http://arxiv.org/abs/1404.0681} {arXiv:1404.0681 [hep-ph]} \BibitemShut
  {NoStop}%
\bibitem [{\citenamefont {Kanemura}\ \emph {et~al.}(2016)\citenamefont
  {Kanemura}, \citenamefont {Kikuchi},\ and\ \citenamefont
  {Yagyu}}]{Kanemura:2015fra}%
  \BibitemOpen
  \bibfield  {author} {\bibinfo {author} {\bibfnamefont {S.}~\bibnamefont
  {Kanemura}}, \bibinfo {author} {\bibfnamefont {M.}~\bibnamefont {Kikuchi}}, \
  and\ \bibinfo {author} {\bibfnamefont {K.}~\bibnamefont {Yagyu}},\ }\href
  {\doibase 10.1016/j.nuclphysb.2016.04.005} {\bibfield  {journal} {\bibinfo
  {journal} {Nucl. Phys.}\ }\textbf {\bibinfo {volume} {B907}},\ \bibinfo
  {pages} {286} (\bibinfo {year} {2016})},\ \Eprint
  {http://arxiv.org/abs/1511.06211} {arXiv:1511.06211 [hep-ph]} \BibitemShut
  {NoStop}%
\bibitem [{\citenamefont {Basler}\ \emph
  {et~al.}(2017{\natexlab{b}})\citenamefont {Basler}, \citenamefont {Krause},
  \citenamefont {Muhlleitner}, \citenamefont {Wittbrodt},\ and\ \citenamefont
  {Wlotzka}}]{Basler:2016obg}%
  \BibitemOpen
  \bibfield  {author} {\bibinfo {author} {\bibfnamefont {P.}~\bibnamefont
  {Basler}}, \bibinfo {author} {\bibfnamefont {M.}~\bibnamefont {Krause}},
  \bibinfo {author} {\bibfnamefont {M.}~\bibnamefont {Muhlleitner}}, \bibinfo
  {author} {\bibfnamefont {J.}~\bibnamefont {Wittbrodt}}, \ and\ \bibinfo
  {author} {\bibfnamefont {A.}~\bibnamefont {Wlotzka}},\ }\href {\doibase
  10.1007/JHEP02(2017)121} {\bibfield  {journal} {\bibinfo  {journal} {JHEP}\
  }\textbf {\bibinfo {volume} {02}},\ \bibinfo {pages} {121} (\bibinfo {year}
  {2017}{\natexlab{b}})},\ \Eprint {http://arxiv.org/abs/1612.04086}
  {arXiv:1612.04086 [hep-ph]} \BibitemShut {NoStop}%
\bibitem [{\citenamefont {Basler}\ \emph
  {et~al.}(2017{\natexlab{c}})\citenamefont {Basler}, \citenamefont
  {Muhlleitner},\ and\ \citenamefont {Wittbrodt}}]{Basler:2017uxn}%
  \BibitemOpen
  \bibfield  {author} {\bibinfo {author} {\bibfnamefont {P.}~\bibnamefont
  {Basler}}, \bibinfo {author} {\bibfnamefont {M.}~\bibnamefont {Muhlleitner}},
  \ and\ \bibinfo {author} {\bibfnamefont {J.}~\bibnamefont {Wittbrodt}},\
  }\href@noop {} {\  (\bibinfo {year} {2017}{\natexlab{c}})},\ \Eprint
  {http://arxiv.org/abs/1711.04097} {arXiv:1711.04097 [hep-ph]} \BibitemShut
  {NoStop}%
\bibitem [{\citenamefont {Kobakhidze}\ and\ \citenamefont
  {Spencer-Smith}(2013)}]{Kobakhidze:2013pya}%
  \BibitemOpen
  \bibfield  {author} {\bibinfo {author} {\bibfnamefont {A.}~\bibnamefont
  {Kobakhidze}}\ and\ \bibinfo {author} {\bibfnamefont {A.}~\bibnamefont
  {Spencer-Smith}},\ }\href {\doibase 10.1007/JHEP08(2013)036} {\bibfield
  {journal} {\bibinfo  {journal} {JHEP}\ }\textbf {\bibinfo {volume} {08}},\
  \bibinfo {pages} {036} (\bibinfo {year} {2013})},\ \Eprint
  {http://arxiv.org/abs/1305.7283} {arXiv:1305.7283 [hep-ph]} \BibitemShut
  {NoStop}%
\bibitem [{\citenamefont {Braathen}\ \emph {et~al.}(2017)\citenamefont
  {Braathen}, \citenamefont {Goodsell},\ and\ \citenamefont
  {Staub}}]{Braathen:2017izn}%
  \BibitemOpen
  \bibfield  {author} {\bibinfo {author} {\bibfnamefont {J.}~\bibnamefont
  {Braathen}}, \bibinfo {author} {\bibfnamefont {M.~D.}\ \bibnamefont
  {Goodsell}}, \ and\ \bibinfo {author} {\bibfnamefont {F.}~\bibnamefont
  {Staub}},\ }\href {\doibase 10.1140/epjc/s10052-017-5303-x} {\bibfield
  {journal} {\bibinfo  {journal} {Eur. Phys. J.}\ }\textbf {\bibinfo {volume}
  {C77}},\ \bibinfo {pages} {757} (\bibinfo {year} {2017})},\ \Eprint
  {http://arxiv.org/abs/1706.05372} {arXiv:1706.05372 [hep-ph]} \BibitemShut
  {NoStop}%
\bibitem [{\citenamefont {Krauss}\ and\ \citenamefont
  {Staub}(2017)}]{Krauss:2017xpj}%
  \BibitemOpen
  \bibfield  {author} {\bibinfo {author} {\bibfnamefont {M.~E.}\ \bibnamefont
  {Krauss}}\ and\ \bibinfo {author} {\bibfnamefont {F.}~\bibnamefont {Staub}},\
  }\href@noop {} {\  (\bibinfo {year} {2017})},\ \Eprint
  {http://arxiv.org/abs/1709.03501} {arXiv:1709.03501 [hep-ph]} \BibitemShut
  {NoStop}%
\bibitem [{\citenamefont {Machacek}\ and\ \citenamefont
  {Vaughn}(1983)}]{Machacek:1983tz}%
  \BibitemOpen
  \bibfield  {author} {\bibinfo {author} {\bibfnamefont {M.~E.}\ \bibnamefont
  {Machacek}}\ and\ \bibinfo {author} {\bibfnamefont {M.~T.}\ \bibnamefont
  {Vaughn}},\ }\href {\doibase 10.1016/0550-3213(83)90610-7} {\bibfield
  {journal} {\bibinfo  {journal} {Nucl. Phys.}\ }\textbf {\bibinfo {volume}
  {B222}},\ \bibinfo {pages} {83} (\bibinfo {year} {1983})}\BibitemShut
  {NoStop}%
\bibitem [{\citenamefont {Machacek}\ and\ \citenamefont
  {Vaughn}(1984)}]{Machacek:1983fi}%
  \BibitemOpen
  \bibfield  {author} {\bibinfo {author} {\bibfnamefont {M.~E.}\ \bibnamefont
  {Machacek}}\ and\ \bibinfo {author} {\bibfnamefont {M.~T.}\ \bibnamefont
  {Vaughn}},\ }\href {\doibase 10.1016/0550-3213(84)90533-9} {\bibfield
  {journal} {\bibinfo  {journal} {Nucl. Phys.}\ }\textbf {\bibinfo {volume}
  {B236}},\ \bibinfo {pages} {221} (\bibinfo {year} {1984})}\BibitemShut
  {NoStop}%
\bibitem [{\citenamefont {Machacek}\ and\ \citenamefont
  {Vaughn}(1985)}]{Machacek:1984zw}%
  \BibitemOpen
  \bibfield  {author} {\bibinfo {author} {\bibfnamefont {M.~E.}\ \bibnamefont
  {Machacek}}\ and\ \bibinfo {author} {\bibfnamefont {M.~T.}\ \bibnamefont
  {Vaughn}},\ }\href {\doibase 10.1016/0550-3213(85)90040-9} {\bibfield
  {journal} {\bibinfo  {journal} {Nucl. Phys.}\ }\textbf {\bibinfo {volume}
  {B249}},\ \bibinfo {pages} {70} (\bibinfo {year} {1985})}\BibitemShut
  {NoStop}%
\bibitem [{\citenamefont {Luo}\ \emph {et~al.}(2003)\citenamefont {Luo},
  \citenamefont {Wang},\ and\ \citenamefont {Xiao}}]{Luo:2002ti}%
  \BibitemOpen
  \bibfield  {author} {\bibinfo {author} {\bibfnamefont {M.-x.}\ \bibnamefont
  {Luo}}, \bibinfo {author} {\bibfnamefont {H.-w.}\ \bibnamefont {Wang}}, \
  and\ \bibinfo {author} {\bibfnamefont {Y.}~\bibnamefont {Xiao}},\ }\href
  {\doibase 10.1103/PhysRevD.67.065019} {\bibfield  {journal} {\bibinfo
  {journal} {Phys. Rev.}\ }\textbf {\bibinfo {volume} {D67}},\ \bibinfo {pages}
  {065019} (\bibinfo {year} {2003})},\ \Eprint
  {http://arxiv.org/abs/hep-ph/0211440} {arXiv:hep-ph/0211440 [hep-ph]}
  \BibitemShut {NoStop}%
\bibitem [{\citenamefont {Staub}(2008)}]{Staub:2008uz}%
  \BibitemOpen
  \bibfield  {author} {\bibinfo {author} {\bibfnamefont {F.}~\bibnamefont
  {Staub}},\ }\href@noop {} {\  (\bibinfo {year} {2008})},\ \Eprint
  {http://arxiv.org/abs/0806.0538} {arXiv:0806.0538 [hep-ph]} \BibitemShut
  {NoStop}%
\bibitem [{\citenamefont {Staub}(2010)}]{Staub:2009bi}%
  \BibitemOpen
  \bibfield  {author} {\bibinfo {author} {\bibfnamefont {F.}~\bibnamefont
  {Staub}},\ }\href {\doibase 10.1016/j.cpc.2010.01.011} {\bibfield  {journal}
  {\bibinfo  {journal} {Comput.Phys.Commun.}\ }\textbf {\bibinfo {volume}
  {181}},\ \bibinfo {pages} {1077} (\bibinfo {year} {2010})},\ \Eprint
  {http://arxiv.org/abs/0909.2863} {arXiv:0909.2863 [hep-ph]} \BibitemShut
  {NoStop}%
\bibitem [{\citenamefont {Staub}(2011)}]{Staub:2010jh}%
  \BibitemOpen
  \bibfield  {author} {\bibinfo {author} {\bibfnamefont {F.}~\bibnamefont
  {Staub}},\ }\href {\doibase 10.1016/j.cpc.2010.11.030} {\bibfield  {journal}
  {\bibinfo  {journal} {Comput.Phys.Commun.}\ }\textbf {\bibinfo {volume}
  {182}},\ \bibinfo {pages} {808} (\bibinfo {year} {2011})},\ \Eprint
  {http://arxiv.org/abs/1002.0840} {arXiv:1002.0840 [hep-ph]} \BibitemShut
  {NoStop}%
\bibitem [{\citenamefont {Staub}(2012)}]{Staub:2012pb}%
  \BibitemOpen
  \bibfield  {author} {\bibinfo {author} {\bibfnamefont {F.}~\bibnamefont
  {Staub}},\ }\href@noop {} {\  (\bibinfo {year} {2012})},\ \Eprint
  {http://arxiv.org/abs/1207.0906} {arXiv:1207.0906 [hep-ph]} \BibitemShut
  {NoStop}%
\bibitem [{\citenamefont {Staub}(2014)}]{Staub:2013tta}%
  \BibitemOpen
  \bibfield  {author} {\bibinfo {author} {\bibfnamefont {F.}~\bibnamefont
  {Staub}},\ }\href {\doibase 10.1016/j.cpc.2014.02.018} {\bibfield  {journal}
  {\bibinfo  {journal} {Comput. Phys. Commun.}\ }\textbf {\bibinfo {volume}
  {185}},\ \bibinfo {pages} {1773} (\bibinfo {year} {2014})},\ \Eprint
  {http://arxiv.org/abs/1309.7223} {arXiv:1309.7223 [hep-ph]} \BibitemShut
  {NoStop}%
\bibitem [{\citenamefont {Staub}(2015)}]{Staub:2015kfa}%
  \BibitemOpen
  \bibfield  {author} {\bibinfo {author} {\bibfnamefont {F.}~\bibnamefont
  {Staub}},\ }\href {\doibase 10.1155/2015/840780} {\bibfield  {journal}
  {\bibinfo  {journal} {Adv. High Energy Phys.}\ }\textbf {\bibinfo {volume}
  {2015}},\ \bibinfo {pages} {840780} (\bibinfo {year} {2015})},\ \Eprint
  {http://arxiv.org/abs/1503.04200} {arXiv:1503.04200 [hep-ph]} \BibitemShut
  {NoStop}%
\bibitem [{\citenamefont {Porod}(2003)}]{Porod:2003um}%
  \BibitemOpen
  \bibfield  {author} {\bibinfo {author} {\bibfnamefont {W.}~\bibnamefont
  {Porod}},\ }\href {\doibase 10.1016/S0010-4655(03)00222-4} {\bibfield
  {journal} {\bibinfo  {journal} {Comput.Phys.Commun.}\ }\textbf {\bibinfo
  {volume} {153}},\ \bibinfo {pages} {275} (\bibinfo {year} {2003})},\ \Eprint
  {http://arxiv.org/abs/hep-ph/0301101} {arXiv:hep-ph/0301101 [hep-ph]}
  \BibitemShut {NoStop}%
\bibitem [{\citenamefont {Porod}\ and\ \citenamefont
  {Staub}(2011)}]{Porod:2011nf}%
  \BibitemOpen
  \bibfield  {author} {\bibinfo {author} {\bibfnamefont {W.}~\bibnamefont
  {Porod}}\ and\ \bibinfo {author} {\bibfnamefont {F.}~\bibnamefont {Staub}},\
  }\href@noop {} {\  (\bibinfo {year} {2011})},\ \Eprint
  {http://arxiv.org/abs/1104.1573} {arXiv:1104.1573 [hep-ph]} \BibitemShut
  {NoStop}%
\bibitem [{\citenamefont {Staub}\ and\ \citenamefont
  {Porod}(2017)}]{Staub:2017jnp}%
  \BibitemOpen
  \bibfield  {author} {\bibinfo {author} {\bibfnamefont {F.}~\bibnamefont
  {Staub}}\ and\ \bibinfo {author} {\bibfnamefont {W.}~\bibnamefont {Porod}},\
  }\href {\doibase 10.1140/epjc/s10052-017-4893-7} {\bibfield  {journal}
  {\bibinfo  {journal} {Eur. Phys. J.}\ }\textbf {\bibinfo {volume} {C77}},\
  \bibinfo {pages} {338} (\bibinfo {year} {2017})},\ \Eprint
  {http://arxiv.org/abs/1703.03267} {arXiv:1703.03267 [hep-ph]} \BibitemShut
  {NoStop}%
\bibitem [{\citenamefont {Martin}(2004)}]{Martin:2003it}%
  \BibitemOpen
  \bibfield  {author} {\bibinfo {author} {\bibfnamefont {S.~P.}\ \bibnamefont
  {Martin}},\ }\href {\doibase 10.1103/PhysRevD.70.016005} {\bibfield
  {journal} {\bibinfo  {journal} {Phys. Rev.}\ }\textbf {\bibinfo {volume}
  {D70}},\ \bibinfo {pages} {016005} (\bibinfo {year} {2004})},\ \Eprint
  {http://arxiv.org/abs/hep-ph/0312092} {arXiv:hep-ph/0312092 [hep-ph]}
  \BibitemShut {NoStop}%
\bibitem [{\citenamefont {Goodsell}\ \emph {et~al.}(2015)\citenamefont
  {Goodsell}, \citenamefont {Nickel},\ and\ \citenamefont
  {Staub}}]{Goodsell:2015ira}%
  \BibitemOpen
  \bibfield  {author} {\bibinfo {author} {\bibfnamefont {M.}~\bibnamefont
  {Goodsell}}, \bibinfo {author} {\bibfnamefont {K.}~\bibnamefont {Nickel}}, \
  and\ \bibinfo {author} {\bibfnamefont {F.}~\bibnamefont {Staub}},\ }\href
  {\doibase 10.1140/epjc/s10052-015-3494-6} {\bibfield  {journal} {\bibinfo
  {journal} {Eur. Phys. J.}\ }\textbf {\bibinfo {volume} {C75}},\ \bibinfo
  {pages} {290} (\bibinfo {year} {2015})},\ \Eprint
  {http://arxiv.org/abs/1503.03098} {arXiv:1503.03098 [hep-ph]} \BibitemShut
  {NoStop}%
\bibitem [{\citenamefont {Braathen}\ and\ \citenamefont
  {Goodsell}(2016)}]{Braathen:2016cqe}%
  \BibitemOpen
  \bibfield  {author} {\bibinfo {author} {\bibfnamefont {J.}~\bibnamefont
  {Braathen}}\ and\ \bibinfo {author} {\bibfnamefont {M.~D.}\ \bibnamefont
  {Goodsell}},\ }\href {\doibase 10.1007/JHEP12(2016)056} {\bibfield  {journal}
  {\bibinfo  {journal} {JHEP}\ }\textbf {\bibinfo {volume} {12}},\ \bibinfo
  {pages} {056} (\bibinfo {year} {2016})},\ \Eprint
  {http://arxiv.org/abs/1609.06977} {arXiv:1609.06977 [hep-ph]} \BibitemShut
  {NoStop}%
\bibitem [{\citenamefont {Martin}(2003)}]{Martin:2003qz}%
  \BibitemOpen
  \bibfield  {author} {\bibinfo {author} {\bibfnamefont {S.~P.}\ \bibnamefont
  {Martin}},\ }\href {\doibase 10.1103/PhysRevD.68.075002} {\bibfield
  {journal} {\bibinfo  {journal} {Phys. Rev.}\ }\textbf {\bibinfo {volume}
  {D68}},\ \bibinfo {pages} {075002} (\bibinfo {year} {2003})},\ \Eprint
  {http://arxiv.org/abs/hep-ph/0307101} {arXiv:hep-ph/0307101 [hep-ph]}
  \BibitemShut {NoStop}%
\bibitem [{\citenamefont {Martin}\ and\ \citenamefont
  {Robertson}(2014)}]{Martin:2014cxa}%
  \BibitemOpen
  \bibfield  {author} {\bibinfo {author} {\bibfnamefont {S.~P.}\ \bibnamefont
  {Martin}}\ and\ \bibinfo {author} {\bibfnamefont {D.~G.}\ \bibnamefont
  {Robertson}},\ }\href {\doibase 10.1103/PhysRevD.90.073010} {\bibfield
  {journal} {\bibinfo  {journal} {Phys. Rev.}\ }\textbf {\bibinfo {volume}
  {D90}},\ \bibinfo {pages} {073010} (\bibinfo {year} {2014})},\ \Eprint
  {http://arxiv.org/abs/1407.4336} {arXiv:1407.4336 [hep-ph]} \BibitemShut
  {NoStop}%
\bibitem [{\citenamefont {Elias-Miro}\ \emph
  {et~al.}(2012{\natexlab{b}})\citenamefont {Elias-Miro}, \citenamefont
  {Espinosa}, \citenamefont {Giudice}, \citenamefont {Isidori}, \citenamefont
  {Riotto},\ and\ \citenamefont {Strumia}}]{EliasMiro:2011aa}%
  \BibitemOpen
  \bibfield  {author} {\bibinfo {author} {\bibfnamefont {J.}~\bibnamefont
  {Elias-Miro}}, \bibinfo {author} {\bibfnamefont {J.~R.}\ \bibnamefont
  {Espinosa}}, \bibinfo {author} {\bibfnamefont {G.~F.}\ \bibnamefont
  {Giudice}}, \bibinfo {author} {\bibfnamefont {G.}~\bibnamefont {Isidori}},
  \bibinfo {author} {\bibfnamefont {A.}~\bibnamefont {Riotto}}, \ and\ \bibinfo
  {author} {\bibfnamefont {A.}~\bibnamefont {Strumia}},\ }\href {\doibase
  10.1016/j.physletb.2012.02.013} {\bibfield  {journal} {\bibinfo  {journal}
  {Phys. Lett.}\ }\textbf {\bibinfo {volume} {B709}},\ \bibinfo {pages} {222}
  (\bibinfo {year} {2012}{\natexlab{b}})},\ \Eprint
  {http://arxiv.org/abs/1112.3022} {arXiv:1112.3022 [hep-ph]} \BibitemShut
  {NoStop}%
\bibitem [{\citenamefont {Bambhaniya}\ \emph {et~al.}(2017)\citenamefont
  {Bambhaniya}, \citenamefont {Bhupal~Dev}, \citenamefont {Goswami},
  \citenamefont {Khan},\ and\ \citenamefont {Rodejohann}}]{Bambhaniya:2016rbb}%
  \BibitemOpen
  \bibfield  {author} {\bibinfo {author} {\bibfnamefont {G.}~\bibnamefont
  {Bambhaniya}}, \bibinfo {author} {\bibfnamefont {P.}~\bibnamefont
  {Bhupal~Dev}}, \bibinfo {author} {\bibfnamefont {S.}~\bibnamefont {Goswami}},
  \bibinfo {author} {\bibfnamefont {S.}~\bibnamefont {Khan}}, \ and\ \bibinfo
  {author} {\bibfnamefont {W.}~\bibnamefont {Rodejohann}},\ }\href {\doibase
  10.1103/PhysRevD.95.095016} {\bibfield  {journal} {\bibinfo  {journal} {Phys.
  Rev.}\ }\textbf {\bibinfo {volume} {D95}},\ \bibinfo {pages} {095016}
  (\bibinfo {year} {2017})},\ \Eprint {http://arxiv.org/abs/1611.03827}
  {arXiv:1611.03827 [hep-ph]} \BibitemShut {NoStop}%
\bibitem [{\citenamefont {Branco}\ \emph {et~al.}(2012)\citenamefont {Branco},
  \citenamefont {Ferreira}, \citenamefont {Lavoura}, \citenamefont {Rebelo},
  \citenamefont {Sher},\ and\ \citenamefont {Silva}}]{Branco:2011iw}%
  \BibitemOpen
  \bibfield  {author} {\bibinfo {author} {\bibfnamefont {G.~C.}\ \bibnamefont
  {Branco}}, \bibinfo {author} {\bibfnamefont {P.~M.}\ \bibnamefont
  {Ferreira}}, \bibinfo {author} {\bibfnamefont {L.}~\bibnamefont {Lavoura}},
  \bibinfo {author} {\bibfnamefont {M.~N.}\ \bibnamefont {Rebelo}}, \bibinfo
  {author} {\bibfnamefont {M.}~\bibnamefont {Sher}}, \ and\ \bibinfo {author}
  {\bibfnamefont {J.~P.}\ \bibnamefont {Silva}},\ }\href {\doibase
  10.1016/j.physrep.2012.02.002} {\bibfield  {journal} {\bibinfo  {journal}
  {Phys. Rept.}\ }\textbf {\bibinfo {volume} {516}},\ \bibinfo {pages} {1}
  (\bibinfo {year} {2012})},\ \Eprint {http://arxiv.org/abs/1106.0034}
  {arXiv:1106.0034 [hep-ph]} \BibitemShut {NoStop}%
\end{thebibliography}%

\end{document}